\documentclass[10pt,a4paper]{article}
\usepackage[left=2.75cm,right=2.75cm,top=2.5cm,bottom=2.5cm]{geometry}
\usepackage[english]{babel}
\usepackage[utf8]{inputenc}

\usepackage[hidelinks]{hyperref}
\usepackage{verbatim}

\usepackage[labelfont=bf]{caption}
\usepackage{subcaption}
\usepackage{color}
\usepackage{colortbl}
\usepackage{amsmath}
\usepackage{amsthm}
\usepackage{amssymb}

\usepackage{pict2e}
\usepackage[final]{pdfpages} 
\usepackage{setspace}
\usepackage{ulem}
\usepackage{placeins}
\usepackage{booktabs}
\usepackage{multirow}
\usepackage{chngpage}
\usepackage{array,multirow,makecell}
\usepackage{sidecap}
\usepackage[ruled,lined,linesnumbered]{algorithm2e}
\sidecaptionvpos{figure}{c}
    \usepackage{thmtools}

\usepackage{chngcntr}
\counterwithout{figure}{section}
\usepackage{pgf,tikz}
\usetikzlibrary{mindmap,trees}
\usetikzlibrary{backgrounds,automata}
\usepackage{tkz-graph}  
\usetikzlibrary[topaths]
\usepgflibrary{arrows.meta} 
\usepgflibrary[arrows.meta] 
\usetikzlibrary{arrows.meta} 
\usetikzlibrary[arrows.meta] 
\usetikzlibrary{shapes.geometric}  
\usetikzlibrary{graphs}
\usetikzlibrary{decorations}
\usetikzlibrary[decorations]
\usetikzlibrary{%
  shapes,
  decorations.shapes,
  decorations.fractals,
  decorations.markings,
	decorations.pathmorphing,
  shadows}

\newcolumntype{C}[1]{>{\centering\arraybackslash }b{#1}}
\usepackage{times} 
\usepackage[title]{appendix} 

\usepackage{chngcntr}
\usepackage{etoolbox}

\usepackage{lipsum}

\usepackage[font={small}]{caption}
 \usepackage{setspace} 
\title{Estimating causal effects of time-dependent exposures on a binary endpoint in a high-dimensional setting}
\author{Vahé Asvatourian$^{1,2}$, Clélia Coutzac$^{3,4}$, Nathalie Chaput$^{3,5}$,\\ Caroline Robert$^{4,6}$, Stefan Michiels$^{1,2}$, Emilie Lanoy$^{1,2}$}
\newcount\mycount
\makeatletter

\newcommand*{\indep}{%
  \mathbin{%
    \mathpalette{\@indep}{}%
  }%
}

\newcommand*{\nindep}{%
  \mathbin{
    \mathpalette{\@indep}{\not}
  }%
}
\newcommand*{\@indep}[2]{%
  \sbox0{$#1\perp\m@th$}
  \sbox2{$#1=$}
  \sbox4{$#1\vcenter{}$}
  \rlap{\copy0}
  \dimen@=\dimexpr\ht2-\ht4-.2pt\relax
  \kern\dimen@
  {#2}%
  \kern\dimen@
  \copy0 
}

\begin{document}
\sloppy 
\maketitle

\begin{flushleft}
$^{1}$University of Paris-Saclay , Univ. Paris-Sud, UVSQ, CESP, INSERM, Villejuif, France;\\
$^{2}$Biostatistics and epidemiology unit, Gustave Roussy, Villejuif, France;\\
$^{3}$Gustave Roussy, Laboratoire d’Immunomonitoring en Oncologie, and CNRS-UMS 3655 and INSERM-US23, Villejuif F-94805, France;\\
$^{4}$Université Paris-Sud, Faculté de Médecine, Le Kremlin Bicêtre, F-94276, France;\\
$^{5}$Université Paris-Sud, Faculté de Pharmacie, Chatenay-Malabry, F-92296, France;\\
$^{6}$Department of Medicine, Dermatology Unit, Gustave Roussy Cancer Campus, Villejuif, France;
\end{flushleft}

\begin{abstract}

Recently, the intervention calculus when the DAG is absent (IDA) method was developed to estimate lower bounds of causal effects from observational high-dimensional data. Originally it was introduced to assess the effect of baseline biomarkers which do not vary over time. However, in many clinical settings, measurements of biomarkers are repeated at fixed time points during treatment exposure and, therefore, this method need to be extended. The purpose of this paper is then to extend the first step of the IDA, the Peter Clarks (PC)-algorithm, to a time-dependent exposure in the context of a binary outcome. 
We generalised the so-called “PC-algorithm” for taking into account the chronological order of repeated measurements of the exposure and propose to apply the IDA with our new version, the chronologically ordered PC-algorithm (COPC-algorithm). The extension includes Firth’s correction. A simulation study has been performed before applying the method for estimating causal effects of time-dependent immunological biomarkers on toxicity, death and progression in patients with metastatic melanoma. 
The simulation study showed that the completed partially directed acyclic graphs (CPDAGs) obtained using COPC-algorithm were structurally closer to the true CPDAG than CPDAGs obtained using PC-algorithm. Also, causal effects were more accurate when they were estimated based on CPDAGs obtained using COPC-algorithm. Moreover, CPDAGs obtained by COPC-algorithm allowed removing non-chronologic arrows with a variable measured at a time t pointing to a variable measured at a time $t’$ where $t’< t$. Bidirected edges were less present in CPDAGs obtained with the COPC-algorithm, supporting the fact that there was less variability in causal effects estimated from these CPDAGs. In the example, a threshold of the per comparison error rate of $0.5\%$ led to the selection of an interpretable set of biomarkers.
Conclusions 
The COPC-algorithm provided CPDAGs that keep the chronological structure present in the data and thus allowed to estimate lower bounds of the causal effect of time-dependent immunological biomarkers on early toxicity, premature death and progression.
\end{abstract}

\section{Background}
The Intervention calculus when the directed acyclic graph (DAG) is absent (IDA) method was recently developed to estimate lower bound of total causal effects from observational data in high-dimensional settings \cite{Maathuis2009}. It was originally introduced to evaluate the effect of time-fixed exposure (gene expression). This method is a combination of Peter Clarks (PC)-algorithm \cite{Spirtes1993} and Pearl's do calculus \cite{Pearl2009e}. The PC-algorithm is a constraint based method for causal structure learning, meaning that it learns the causal structure based on the conditional dependencies of the observational distribution. The output of the PC-algorithm results in a CPDAG (completed partially DAG) that encodes conditional dependencies of the data in a class of DAGs (Directed acyclic graphs) called \textit{Markov Equivalent}. Then, based on the DAGs in the \textit{Markov Equivalence Class}, causal effects are estimated using Pearl’s do calculus \cite{Pearl2009e} (see section Computation of the causal effects). However, in many clinical settings, time-dependent biomarker values under treatment or changes in biomarkers from baseline are of interest. If the true DAG was known, the commonly used marginal structural model (MSM) approach could applied to estimate causal effects in the case of time-dependent covariates and outcome \cite{Angel2000,Robins2000a}. In our setting, the true DAG being unknown, causal effects could not be identified using MSM.\\

In the 2010s, new anti-cancer treatments targeting immune checkpoints have been introduced: the wave of these immunotherapies began with the anti CTLA-4 treatment which showed a survival benefit in patients with metastatic melanoma \cite{Robert2011,Hodi2010}. More recently, promising results in lung and kidney cancers have also been obtained \cite{Drake2014}. Nevertheless, only a subgroup of patients seem to benefit from this treatment: about 20$\%$ of patients with metastatic melanoma  treated with ipilimumab were long-term survivors (3 years) \cite{Schadendorf2015}. Moreover, immune related toxicity such as colitis occurs in 8 to 22$\%$ of treated patients \cite{Garbe2011a}. The goal of immunotherapy is to amplify the immune system response against cancer cells. Thus, one can observe the evolution of the treatment by looking at the immune system. Predictive and/or prognostic markers are ideally validated through clinical trials including randomized studies, which are the gold standard \cite{Sargent2005,Buyse2011}. Before being evaluated in randomized trial, candidate immunological biomarkers can be identified from high-dimensional data, collected in an observational or non-randomized setting.\\

Our objective was to develop methods to identify the causal effects of time-dependent exposures on a binary endpoint in a high-dimensional setting, with an application of time-dependent immunological biomarkers in a non-randomized prospective study in oncology. . However, the PC-algorithm has never been applied on data measured repeatedly at a fixed time points, and the chronological order among data is not respected when using PC-algorithm. The first step was then to find the true CPDAG by extending the PC-algorithm to chronologically ordered measures and then to estimate robust causal effects based on the CPDAG estimated using our version of the PC-algorithm. To ensure the accuracy and the efficiency of our method, we made a simulation study where we compared the CPDAGs’ structure obtained using PC-algorithm and our method. Then we compared the estimation of true causal effects calculated based on CPDAGs obtained from both methods. Due to collinearity among time-dependent biomarkers, we added for the first time the Firth’s correction while estimating causal effects to avoid instability of the maximum likelihood estimates. After the simulation study, we applied both PC-algorithm and our method to real dataset of time-dependent immunological biomarkers.

\section{Material and methods}
\subsection{Graph definitions and notations}
Let $G(N,E)$ be a graph consisting of nodes $N$ and edges $E$. Nodes represent random variables $N=\{X_{1},…,X_{p}\}$ and edges represent the links between them. An edge can be either directed $X_{i}\rightarrow X_{j}$(in this case, $X_{i}$ is a parent of $X_{j}$ and $X_{j}$ is a descendant of $X_{i}$) or undirected $X_{i} - X_{j}$. A graph with only undirected edges is said to be an undirected graph whereas a directed graph is made of only directed edges. A partially directed graph contains both directed and undirected edges. 
Two nodes are said to be adjacent if they are connected by an edge (either directed or undirected). A path is a sequence of nodes in which all pairs are adjacent. A path can be either open or closed. A path is open when there is no collision between two arrows pointing to the same node on the path (i.e. the path from  $X_{i}$ to  $X_{m}$ in (\ref{eq1}) is open). 

\begin{equation}\label{eq1}
X_{i}\rightarrow X_{j}\rightarrow X_{k} \rightarrow X_{m} \leftarrow X_{l}
\end{equation}

A path is closed when there is a collision between two arrows which point to the same node of the path, this variable is a collider (i.e, the path from $X_{i}$ to $X_{i}$  in (\ref{eq1}) is closed). We denote $X_{k}$ as a descendent of $X_{i}$ (and $X_{i}$ an ancestor of $X_{k}$) if there is a path that starts from $X_{i}$ and ends to $X_{k}$ by following the direction of the arrows (\ref{eq1}). We also denote $pa(X_{i},G)$ as the parents of $X_{i}$  in $G$ by the set of variables pointing to $X_{i}$. A graph is called acyclic when no path starts and ends at the same node. A graph which is acyclic and has directed edges is called a directed acyclic graph (DAG). A DAG is complete or statistical when all pairs of nodes are adjacent, whereas a DAG is causal when all common causes of any variable are on the graph, i.e. any parent is a cause of its descendants. Therefore, a causal DAG is informative whereas a complete DAG is non-informative because a lack of arrow means an absence of a direct causal effect.\\

A graph encodes (conditional) independence relationships through the concept of d-separation
\cite{Pearl1995}. If two nodes are d-separated by a set of nodes, then the variables corresponding to
the nodes are conditionally independent given this set of variables. The set of these given variables
is then called separation set S. Multiple DAGs can be compatible with a same set of underlying
conditional independences. Let a skeleton be the graph obtained by removing all arrowheads from the DAG and the \textit{v-structures} a subgraph of 3 nodes filling two conditions: 1) both arrows are not pointed on  $X_{j}$($X_{j}$  is not a collider) and 2) where $X_{i}$ and $X_{j}$ are not adjacent.\\  
The DAGs $X_{i} \rightarrow X_{j} \rightarrow X_{k}$, $X_{i} \leftarrow X_{j} \leftarrow X_{k}$ and $X_{i}\leftarrow X_{j}\rightarrow X_{k}$ in which the two conditions hold belong to the same \textit{Markov equivalence class} and are called \textit{Markov equivalent}. A whole equivalence class can be summarised in a graph that has the same skeleton and includes the directed arrows of all DAGs in the equivalence class. Edges which are directed differently across the DAGs in the equivalence class are represented with bidirected arrows (or simply edges). This graph with both undirected and directed edges is called a Completed Partially DAG (CPDAG).\\

\subsection{Causal effect estimation in high dimensional settings}
\subsubsection{IDA}
\begin{figure}[ht]

\begin{center}
\includegraphics[scale=0.30]{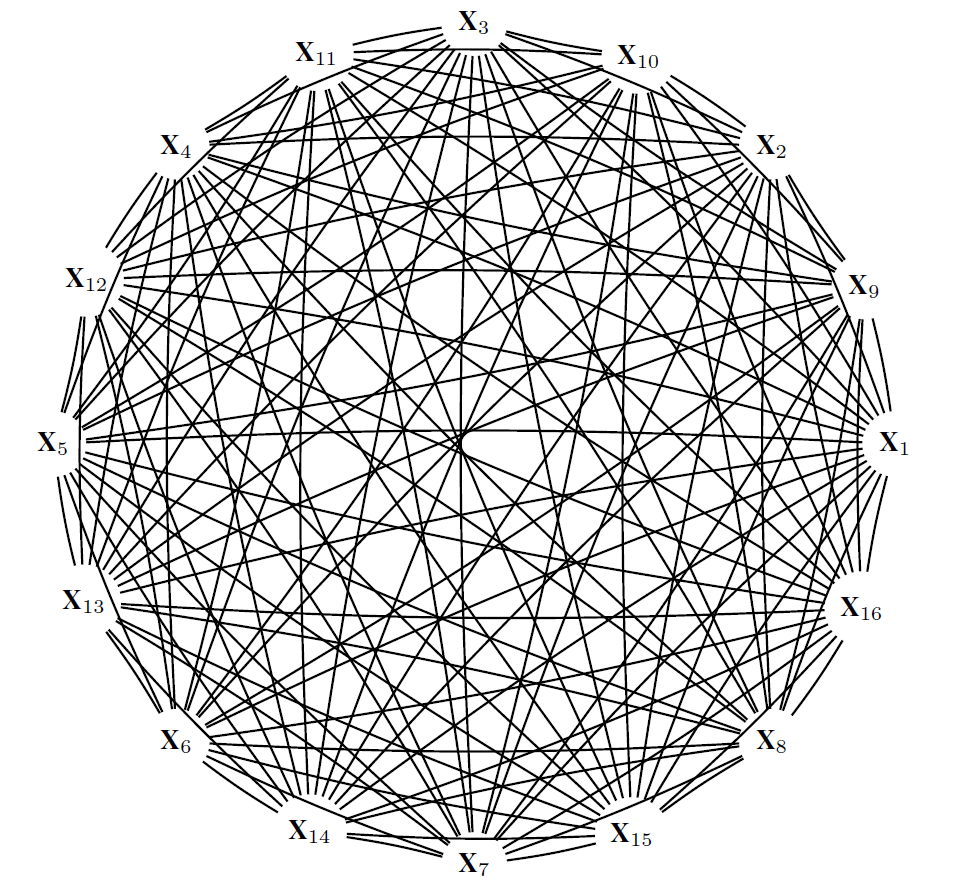}
\caption{A complete undirected graph.\label{figure1}}
\end{center}

\end{figure}

When the relationships between variables are not oriented, the DAG cannot be identified. With many variables in a high-dimensional setting, it is not possible to determine which nodes are ancestors and which are descendants. The only possible initial graph that can be drawn based on high-dimensional data is a complete undirected graph which is non-informative as in Figure \ref{figure1}. The intervention calculus when the DAG is absent (IDA) method has been introduced to determine the CPDAG from the observational data and to estimate lower bounds of the absolute values of the total causal effects in the case where all variables (including outcome) are continuous \cite{Maathuis2009}; and has been extended to the case where all variables are binary \cite{Kalisch2010a}. The first objective of the IDA is to estimate the CPDAG and its Markov equivalence class that contain the true causal DAG from the observational data by using a causal learning algorithm such as the PC-algorithm \cite{Spirtes1993}. Then the intervention calculus \cite{Pearl2009e,Pearl2003} is used on  the $m$ DAGs$_{j}$ of the \textit{Markov equivalence class} $j=1,…,m,$ to estimate the $p\times m$ matrix $\theta$ of causal effects $\theta_{ij}$ of each covariate $X_{i} (i=1,…,p)$ on $Y$.  

However, estimating the true causal effect is impossible when a unique DAG is not identifiable. To determine whether or not a covariate has a potential causal effect, the minimum absolute causal effect of a covariate is defined as $\hat{\beta}_{i}=min_{j}(|\hat{\theta}_{i,j}|)$. Then a ranking of covariates' causal effects is made based on these lower bounds, where $\beta_{i1}$ is the lower bound of the covariate i with the rank 1:

\begin{equation} \label{eq2}
\hat{\beta}_{i1} \geq \hat{\beta}_{i2} \geq \hdots \geq \hat{\beta}_{ip}.
\end{equation}

Determining all the DAGs that are present in the \textit{Markov equivalence class} can be highly computationally intensive in a high-dimensional setting. Nevertheless, rather than computing all the DAGs, it is still possible to determine the set of parents used for adjusting by extracting them from the CPDAG. The local algorithm used by Maathuis et al. \cite{Maathuis2009} checks if the parents are locally valid (if they create or not a new collider) in the CPDAG and all causal estimates for a single covariate $X_{i}$ on $Y$ are in the multiset $\theta_{i}=\{ \theta_{ij}\}$ with $j \in \{1,\hdots,m\}$ and $i \in \{1 \hdots, p\}$. Contrary to a set, in a multiset the replication of an element matters. For instance, the multisets $\{a,a,b\}$ and $\{a,b\}$ are not equal while the sets $\{a,a,b\}$ and $\{a,b\}$ are. The multiset allows the multiplicity of an element. Finally, the assumptions made in the IDA are:

\begin{enumerate}
\item There are no hidden variables.
\item The joint distribution of covariates $X_{i},\hdots,X_{p}$  is normal and faithful to the true (unknown) DAG.
\item Covariates$X_{i},\hdots,X_{p}$ have equal variance.
\end{enumerate}
The IDA method developed by Maathuis et al is implemented in the R-package pcalg \cite{Kalisch2012}.

 \subsubsection{PC-algorithm} 

The PC-algorithm is a constraint based method for causal structure learning \cite{Spirtes1993,Maathuis2015}, meaning  that it learns the causal structure based on the conditional dependencies between variables. A sketch of the PC-algorithm is given in algorithm \ref{PC}.

\begin{algorithm}

 \KwIn{Data $\mathcal{D}=\{X_{1},...,X_{p}\}$, set of ordered vertex $\textbf{V}$, significance parameter $\alpha$} 
 
Determine the skeleton\;
Determine the \textit{v-structures}\;
Orient as many of the remaining edges as possible\;
	
	  \KwOut{CPDAG $\hat{G}$  }
 
\caption{PC-algorithm \label{PC}}

\end{algorithm}

First, it estimates the skeleton of the underlying structure by checking all given conditional dependencies between each variable at a significance level $\alpha$. If no information on dependencies is given, then the graph used as input is an undirected graph such as in Figure \ref{figure1}. Once the skeleton is obtained, edges are oriented in the v-structures to meet the conditional dependencies and finally the CPDAG is obtained by directing as many remaining edges as possible according to three rules \cite{Meek1995a}:

\begin{enumerate}
\item When there is a triple $X_{i}\rightarrow X_{j} - X_{k}$ and $X_{i}$, $X_{k}$ not adjacent, orient $X_{j} - X_{k}$ as $X_{j}\rightarrow X_{k}$  

\item When there is a triple $X_{i} \rightarrow X_{k} \rightarrow X_{j}$, orient $X_{i} - X_{j}$ as $X_{i}\rightarrow  X_{j}$   
\item When there are two triples $X_{i} - X_{l} \rightarrow X_{j}$ and 
$X_{i} - X_{k} \rightarrow X_{j}$ with $X_{k}$ and $X_{l}$ not adjacent, $X_{i} - X_{j}$ is oriented into $X_{i} \rightarrow X_{j}$.
\end{enumerate}

Even though the PC-algorithm has been shown to be consistent in high-dimensional settings \cite{Kalisch2007a}, one of its issues remains the effect of the set of ordered variables O in the final output. In fact, the order of the variables determines which pair of nodes is tested first, determining which edges are removed first and so affecting which tests are considered later on. This order dependence impacts robustness of the results in high-dimensional settings. Two different solutions have been suggested: the stability ranking and the PC-stable, which will be outlined below. Before running the algorithm, the multiple testing requires to specify the significance level (cut-off) $\alpha$ for the conditional independence tests. In fact, setting $\alpha$ to a certain value means that only conditional dependencies with a p-value under $\alpha$ are kept. Thus, running PC-algorithm with a small value of alpha leads to obtain sparser graphs. 

\subsubsection{Stability selection}

To deal with the order dependence issue of the PC-algorithm in the IDA which can lead to poor robustness, Stekhoven et al. proposed to add a stability selection step \cite{Meinshausen2010} to IDA. This method, called Causal Stability ranking (Cstar) \cite{Stekhoven2012}, is based on a re-sampling approach. The IDA is run over 100 independent random subsamples and then in each subsampling run, the variables are ranked according to (\ref{eq2}). At the end of all runs, the relative frequencies $\pi_{i}$ of covariates appearing among the top of $q$ variables are used to define a stable ranking:

\begin{equation}
\hat{\Pi}_{i1} \geq \hat{\Pi}_{i2} \geq \hdots \geq \hat{\Pi}_{ip}.
\end{equation}

For a given $q$, a bound for the per-comparison error rate (PCER) which can be seen as the false positive error rate is given by:

\begin{equation}
\frac{1}{2 \hat{\Pi}_{j}-1} \frac{q^{2}}{p^{2}}
\end{equation}

\subsubsection{PC-stable}
Another approach that considers the order dependence issue of the PC-algorithm was explored by Colombo and Maathuis by introducing an order independent version of the PC-algorithm called PC-stable \cite{Colombo2014}. In step 1 of the PC-stable version,  the adjacency set of  all variables are stored after each change in the size of the separation set (see section 4.1 of \cite{Colombo2014}); removing an edge will not affect which conditional independencies are checked for other pairs of variables. In addition, they also showed that the combination of the stability selection with PC-stable in gave more reliable edges than PC-stable alone on  yeast gene expression data \cite{Colombo2014}. 
\subsection{Extension to a time-dependant exposure}

We aimed to extend the IDA by integrating time-dependent exposures in the PC-stable step. Based on chronologically ordered data, the resulting CPDAG should not contain arrow from a descendant to a parent $X_{1,t'} \rightarrow X_{1,t}$ where $t<t'$ since the value of a variable at time $t'$ cannot influence a past value of the same variable. This means also that in the first step, when looking at conditional dependencies between two variables measured at time $t$ and $t^{*}$ where $t \geq t^{*}$, variables measured at a time $t'$ where $t'> t$ and  $t'>t^{*}$ should not be tested for the separation set S. \\

\begin{figure}[ht]

\begin{center}
\includegraphics[scale=0.30]{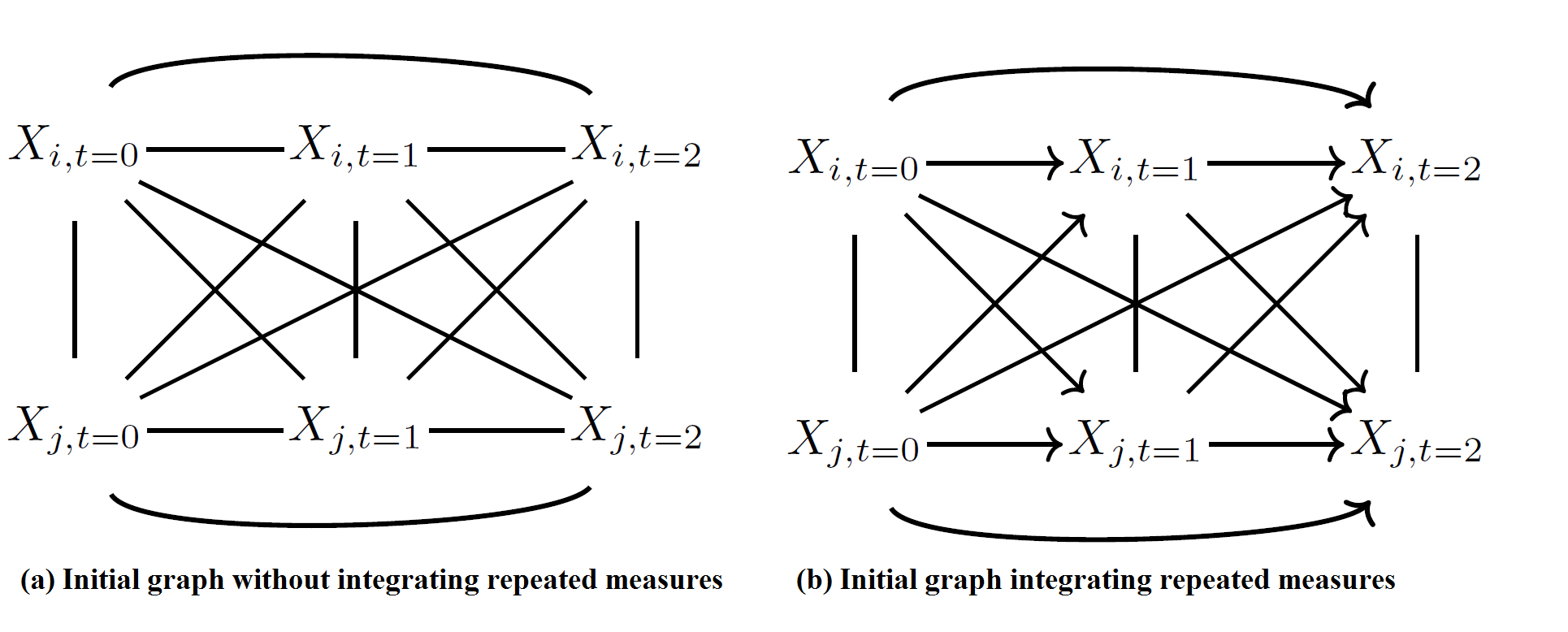}
\caption{Initial graphs used as input for the IDA with and without chronological a priori information for 2 variables $X_{i},X_{j}$ measured at 3 time points $t_{1},t_{2}$  and $t_{3}$. \label{figure2}
}
\end{center}

\end{figure}

This can be done by adding chronological order information among the variables in addition to the conditional independence information as input of the PC-stable algorithm. The result of combining these two types of information can be viewed as a partially directed graph. In the partially directed graph, all edges between variables measured at different times should be directed chronologically, from parents to descendants, and edges between variables measured at the same time remained undirected. Differences between the two initial graphs with 2 variables measured at 3 time points are shown in Figure \ref{figure2}. A global sketch of the chronologically ordered PC-stable is shown in algorithm \ref{COPC}.

\begin{algorithm}

 \KwIn{Data $\mathcal{D}=\{X_{1},...,X_{p}\}$, set of ordered vertex $\textbf{V}$, significance parameter $\alpha$, chronological information} 
 
Determine the skeleton (using chronological information)\;
Orient edges according to the chronological information\;
Determine the \textit{v-structures}\;
Orient as many of the remaining edges as possible\;	
	  \KwOut{CPDAG $\hat{G}$  } 
\caption{Chronologically ordered PC-algorithm \label{COPC}}

\end{algorithm}
The modified step 1 leads thus to determine a skeleton at the end of step 1 while testing only conditional independences within a same time slice. Then before determining the \textit{v-structures}, the chronological information is used in step 2 to orient $X_{i,t} - X_{i,t+1}$ into $X_{i,t} \rightarrow X_{i,t+1}$. We will call this extension of PC-algorithm chronologically ordered PC-stable (COPC-stable) when using the order dependent version or the COPC- algorithm when not.

\subsubsection{Estimation of the causal effect of repeated continuous covariates on a binary outcome}

The estimation of causal effects for data with only continuous or only discrete data has been largely discussed \cite{Pearl2009e,Hernan2018}. In estimate causal effect of repeated continuous covariates on a binary outcome, the collinearity may address an issue of unstable maximum likelihood estimates. Therefore we used the Firth’s correction to address this problem \cite{Firth1993,Heinze2002}. Our model is detailed in the appendix A. 

\subsubsection{Simulations}

To compare our algorithm COPC to the PC-stable algorithm, we used simulations. We generated random weighted DAGs with a given number of variables $p$ per visit, a given number of visits $n_{visits}$(corresponding to measurements of these variables) and a single binary outcome. To simulate collinearity between repeated measures, we generated the repeated covariates data from a multivariate distribution that uses an autoregressive model for the correlation between biomarkers:

\begin{equation}
X \sim N\Bigg(\mu = \begin{pmatrix} \mu_{1}\\ \vdots \\ \mu_{p}\end{pmatrix}, \Sigma=
\begin{pmatrix}
\rho^{0}\sigma^{2}& \hdots & 	\rho^{n_{visits}}\sigma^{2}\\
\vdots & \ddots & \vdots \\
\rho^{n_{visits}}\sigma^{2} & \hdots & \rho^{0}\sigma^{2}
\end{pmatrix}\Bigg),
\end{equation}
where $\rho$ is the correlation between biomarkers. We choose to set $\mu=0$,  $\sigma^{2}=1$ and vary $\rho$ from 0.5 to 0.7. We also tried different number of visits and observations from 3 to 6 and 50 to 1000 respectively. To evaluate the two methods, we compared the capacity of recovering the true CPDAG through the sensibility and the specificity which determine the capacity of detecting the true presence of an arrow and the true absence of an arrow respectively. We also calculated the Structural Hamming distance (SHD) described by Tsamardinos \cite{Tsamardinos2006} which is a score to evaluate the structural distance from an estimated graph to a true graph. The SHD was calculated as follows: SHD was incremented when there was a wrong connection (i.e. there was an arrow in the estimated CPDAG that was absent in the true CPDAG), and a missed edge (i.e. there was no arrow in the estimated CPDAG that was present in the true CPDAG). The accuracy of the causal effects estimation was explored by calculating the mean squared errors (MSE). The full details of the simulations set-up are available in appendix B. 

\subsection{Application}
The method described above was applied on observational data of repeated immunological biomarkers from patients treated with ipilimumab for metastatic melanoma. The objective was to highlight immunologic biomarkers that had a causal effect on early toxicity, premature death and progression. 

\subsubsection{Patients}
Patients with metastatic melanoma treated with ipilimumab were prospectively enrolled at the Gustave Roussy Cancer Campus. Ipilimumab was administered intravenously every 3 weeks. Immunological biomarkers were measured at each visit prior each ipilimumab infusion $(V_{1}, V_{2}, V_{3}, and V_{4})$.
\subsubsection{Outcomes}
Three binary outcomes such as toxicity, premature death and progression were investigated. Early toxicity was defined as occurrence of colitis 12 weeks after treatment initiation. Premature death referred at death 12 weeks after treatment initiation. Progression was defined as an increase of at least 20$\%$ in tumor size or occurrence of new lesions 6 months after treatment initiation.  
\subsubsection{Imunological biomarkers}
Several biological models were used representing different level of immunological expression (Table \ref{tab:biomod}). Model 1 represents adaptive T cells in a global way while model 3 represents subgroup of adaptive T cells. In all three models biomarkers with a potentially known effect were incorporated. For convenience, all biomarkers have been anonymised in the main text of this page but are fully detailed in appendix C.   

\begin{table}[h]
\centering
\caption{Biological models representing different level of immunological expression \label{tab:biomod}}

\begin{tabular}{cccc}
\hline
Model & Common biomarkers (n) & Adaptative T cells (n) & \begin{tabular}[c]{@{}c@{}}Number total \\ of covariates\end{tabular} \\ \hline
1 & \begin{tabular}[c]{@{}c@{}}Non-immunologic and innate  \\  immunological biomarkers
(29)\end{tabular} & CD4 and CD8 (8) & 37 \\ 
2 & \begin{tabular}[c]{@{}c@{}}Non-immunologic and innate  \\  immunological biomarkers
(29)\end{tabular} & \begin{tabular}[c]{@{}c@{}}CD4/CD8 expressing polarization\\ and domiciliation markers(148)\end{tabular} & 177 \\ 
3 & \begin{tabular}[c]{@{}c@{}}Non-immunologic and innate  \\  immunological biomarkers
(29)\end{tabular} & \begin{tabular}[c]{@{}c@{}}Subgroup of CD4 and CD8 \\ expressing polarization \\ and domiciliation markers (232)\end{tabular} & 261 \\ \hline
\end{tabular}
\end{table}

\subsubsection{Representation}
To identify the dependency structure of the data, CPDAGs were estimated using the PC-algorithm. To resume the (conditional) dependencies present in all CPDAGs, Kalisch et al \cite{Kalisch2010a} proposed to aggregate edges in a present in CPDAGs from a resampled dataset rather than showing a single estimate of the CPDAG. Only edges present in $20\%$ of the CPDAGs are drawn and their thickness is proportional to the number of CPDAGs in which the edge was present.  

\subsubsection{Missing data}
In our melanoma application, around $15\%$ of missing data were imputed using multivariate imputation by chained equations (MICE) \cite{Buuren2011}. Missingness graphs \cite{Mohan2013} are substantives tools that have been developed to study the missingness mechanisms and the recoverability of a missing variable. We applied missingness graphs on our data in order in to identify the missingness mechanisms and the recoverability. In missing at random (MAR) case, the missing values can be recovered without bias; while in the missing not at random (MNAR) case, the missing values could be recovered with some little bias. Full details are provided in the appendix D.

\section{Results}
\subsection{Simulations}
The results of the simulations are presented in table 2.

\begin{table}[h]
\centering
\caption{Average sensibility, specificity and SHD according PC-stable and COPC-stable over 500 replicates simulated based on 2 DAGs with different number of visits.\label{table2}}

\scalebox{0.85}{
\begin{tabular}{ccccccccc}
\hline
\textbf{$n_{visits}$} & \textbf{nobs} & \textbf{alpha} & \textbf{\begin{tabular}[c]{@{}c@{}}Se PC-stable\\  (sd) \%\end{tabular}} & \textbf{\begin{tabular}[c]{@{}c@{}}Se COPC-\\ stable (sd) \%\end{tabular}} & \textbf{\begin{tabular}[c]{@{}c@{}}Sp PC-stable\\  (sd) \%\end{tabular}} & \textbf{\begin{tabular}[c]{@{}c@{}}Sp COPC-\\ stable (sd) \%\end{tabular}} & \textbf{\begin{tabular}[c]{@{}c@{}}SHD PC-\\ stable (sd)\end{tabular}} & \textbf{\begin{tabular}[c]{@{}c@{}}SHD COPC-\\ stable (sd)\end{tabular}} \\ \hline
\multirow{4}{*}{4} & \multirow{2}{*}{1000} & 0.02 & 58.1(0.6) & 63.2(0.5) & 98.7(0.1) & 98.8(0.1) & 333 (9) & 279 (7) \\ \cline{3-9} 
 &  & 0.2 & 58.1(0.5) & 64.1(0.5) & 98.6(0.1) & 98.5(0.1) & 340 (9) & 288 (8) \\ \cline{2-9} 
 & \multirow{2}{*}{50} & 0.02 & 54.9(0.5) & 57.0(0.6) & 99.2(0.1) & 99.4(0.1) & 338 (8) & 299 (9) \\ \cline{3-9} 
 &  & 0.2 & 56.3(0.6) & 59.0(0.6) & 98.9(0.1) & 99.1(0.1) & 339 (9) & 296 (8) \\ \hline
\multirow{4}{*}{6} & \multirow{2}{*}{1000} & 0.02 & 56.6(0.4) & 60.7(0.4) & 99.0(0.1) & 98.9(0.1) & 504 (10) & 466 (8) \\ \cline{3-9} 
 &  & 0.2 & 56.6(0.4) & 61.6(0.4) & 98.9(0.1) & 98.6(0.1) & 521 (9) & 491 (10) \\ \cline{2-9} 
 & \multirow{2}{*}{50} & 0.02 & 54.1(0.3) & 55.6(0.4) & 99.4(0.1) & 99.5(0.1) & 484 (9) & 455 (11) \\ \cline{3-9} 
 &  & 0.2 & 55.2(0.4) & 57.5(0.5) & 99.2(0.1) & 99.3(0.1) & 494 (11) & 454 (12) \\ \hline
\end{tabular}
}
\end{table}

Overall, COPC-stable outperformed PC-stable in terms of sensibility, meaning that the percentage of false positive was lower in the CPDAGs estimated with COPC-stable rather than the CPDAGs estimated with PC-stable. In terms of specificity, both algorithms showed excellent results. In scenarios with a greater alpha level regarding other parameters, sensibility rose while specificity decreased. Reducing the number of observations from 1000 to 50 made the sensitivity and specificity slightly underestimated.\\ 
The COPC-stable SHD was lower than the PC-stable in all scenarios, meaning that, as compared with  CPDAGs estimated with PC-stable, CPDAGs estimated with COPC-stable had a structure closer to the true CPDAG  (see table 2). \\
In terms of accuracy, the estimations of causal effects based on CPDAGs estimated with COPC-stable were more accurate than the ones using CPDAGs estimated with PC-stable (see appendix E for results of all scenarios).

\subsection{Application}

Both IDA and our extension have been applied on our observational data of repeated immunological biomarkers from patients treated with immunotherapy for metastatic melanoma. They have been repeatedly run 300 times on subsamples of size n=30. The tuning parameter $\alpha$ was set to 0.02.\\ 
As expected, CPDAGs obtained using a naïve PC-stable from unordered repeated measures led to non-chronological ordered paths in all three models (Figure 3) as compared with paths identified through COPC.

\begin{table}[h]
\centering
\caption{Average number of edges (standard deviation) in the CPDAG according to the version of the PC-algorithm and the model over 300 runs with $\alpha=0.02$.\label{tab:aretedirigees}}

\begin{tabular}{@{}ccccc@{}}
\toprule
\textbf{} & \textbf{Directed edges (sd)} & \textbf{Bidirected edges (sd)} & \begin{tabular}[c]{@{}l@{}}\textbf{Non-chronologically } \\ \textbf{ ordered edges (sd) }\end{tabular}&\textbf{Total (sd)} \\ \midrule
\textbf{PC-algo (model 1)} & 0 (0.1) & 23 (0.2)&17 (0.2) & 22 (0.1) \\
\textbf{COPC-algo (model 1)} & 27 (0.2) & 7 (0.3) &0 (0)& 34 (0.3) \\
\textbf{PC-algo(model 2) } & 3 (0.2) & 120 (0.5) & 76 (0.5)&122 (0.5) \\
\textbf{COPC-algo (model 2)} & 120 (0.6) & 58 (0.4) & 0 (0) & 178 (0.6)\\ 
\textbf{PC-algo (model 3)} & 5 (0.2) & 197 (0.8) & 101 (0.7)&202 (0.7) \\
\textbf{COPC-algo (model 3)} & 153 (1) & 112 (0.7) & 0 (0)& 265 (1) \\ \bottomrule
\end{tabular}
\end{table}

\begin{figure}[h!]

\begin{center}
\includegraphics[scale=0.35]{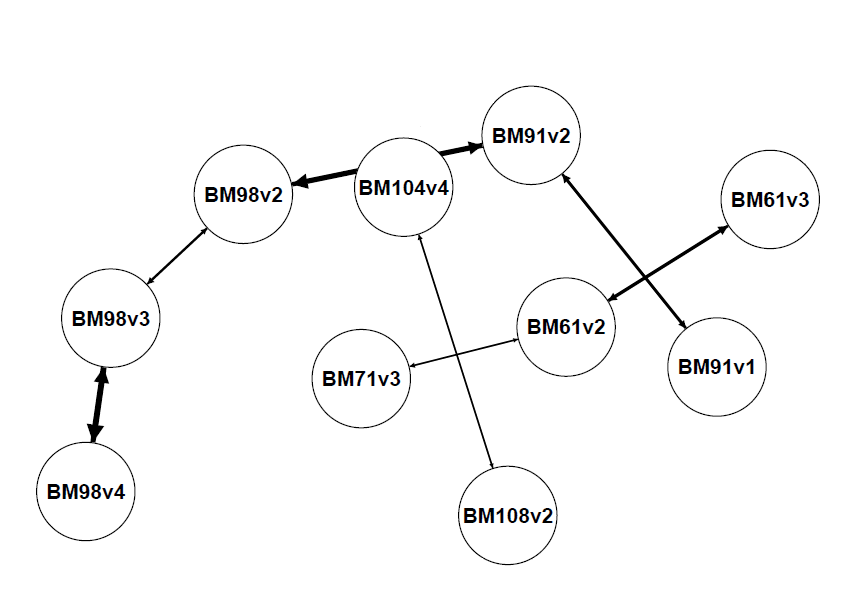}
\caption{Subset of the summary CPDAGs (Completed partially DAGs) of the model 3 
in the metastatic melanoma example using naive PC-stable over 300 runs. Only edges with a frequency $> 0.20$ are present. The thickness of edges is proportional to their frequency.\label{figure3}}
\end{center}

\end{figure}

Table \ref{tab:aretedirigees} shows the average number of edges according to the version of the PC-algorithm and the model. As compared with PC-stable, the percentage of bidirected edges among all edges  using COPC-stable was on average smaller in all three models, $100\%$ vs $28\%$ for model 1, $98\%$ vs $40\%$ for model 2 and $97\%$ vs $52\%$ for model 3. Moreover, Table 3 shows how many edges are defined wrongly into the final CPDAG. For instance, in model 1, when using a naïve approach of the PC-stable, the resulting CPDAG had on average 14 bidirected edges that were between two variables measured at different times. When looking at Table 4, the number of values in each multiset $\theta_{i}$ also called \textit{ambiguity} ($\hat{a}$) of the multiset \cite{Maathuis2009} was smaller when using COPC-stable rather than PC-stable for a same value of alpha $(\alpha=0.02)$. The maximum ambiguity reached in our application was 3.

\begin{table}[h!]
\centering
\caption{Probability of having a certain ambiguity $\hat{a}$ for biomarkers with an alpha level at 0.02 according to the version of the PC-algorithm (PC-Stable/ COPC-stable) over  300 with $\alpha=0.02$\label{tab:ambiguite}}

\begin{tabular}{ccccccc}
\hline
\multirow{2}{*}{\textit{Ambiguity}} & \multicolumn{2}{l}{Model 1} & \multicolumn{2}{l}{Model 2} & \multicolumn{2}{l}{Model 3} \\ \cline{2-7} 
 & PC-stable & COPC-stable & PC-stable & COPC-stable & PC-stable & COPC-stable \\ \hline
$\hat{a}=1$ & 0.243 & 0.676 & 0.153 & 0.599 & 0.061 & 0.437 \\ 
$\hat{a}=2$ & 0.568 & 0.297 & 0.655 & 0.356 & 0.693 & 0.494 \\ 
$\hat{a}=3$ & 0.189 & 0.027 & 0.192 & 0.045 & 0.245 & 0.069 \\ \hline
\end{tabular}
\end{table} 

\subsubsection{Estimating time-dependent causal effects in the melanoma example}

After estimating the CPDAG using COPC-stable, causal effects were estimated using Pearl’s do-calculus.  To determine which biomarker had a robust causal effect, we intended to select biomarkers with PCER threshold $\leq 0.5\%$. In model 1, there were no biomarkers with a PCER < 0.005. Figures 4 and 5 show histograms of causal effects on our three outcomes death, progression and toxicity based on model 2 and 3. The causal effects seem almost uniformly distributed between 0 and 1 in our example for models 2 and 3. However, immunological biomarkers with a PCER under $0.5\%$ had a causal effect concentrated between 0.6 and 0.8 for models 2 and 3 for all outcomes. On the other hand, causal effects sizes of immunological biomarker with PCER $>0.5\%$ were spread in a wide range from 0 to 1.\\ 
 Tables 5 shows the top effect biomarkers among those selected for models 2 (see appendix F for the list of all selected immunological biomarkers).

\begin{figure}[h!]
\centering
   
        \includegraphics[scale=0.60]{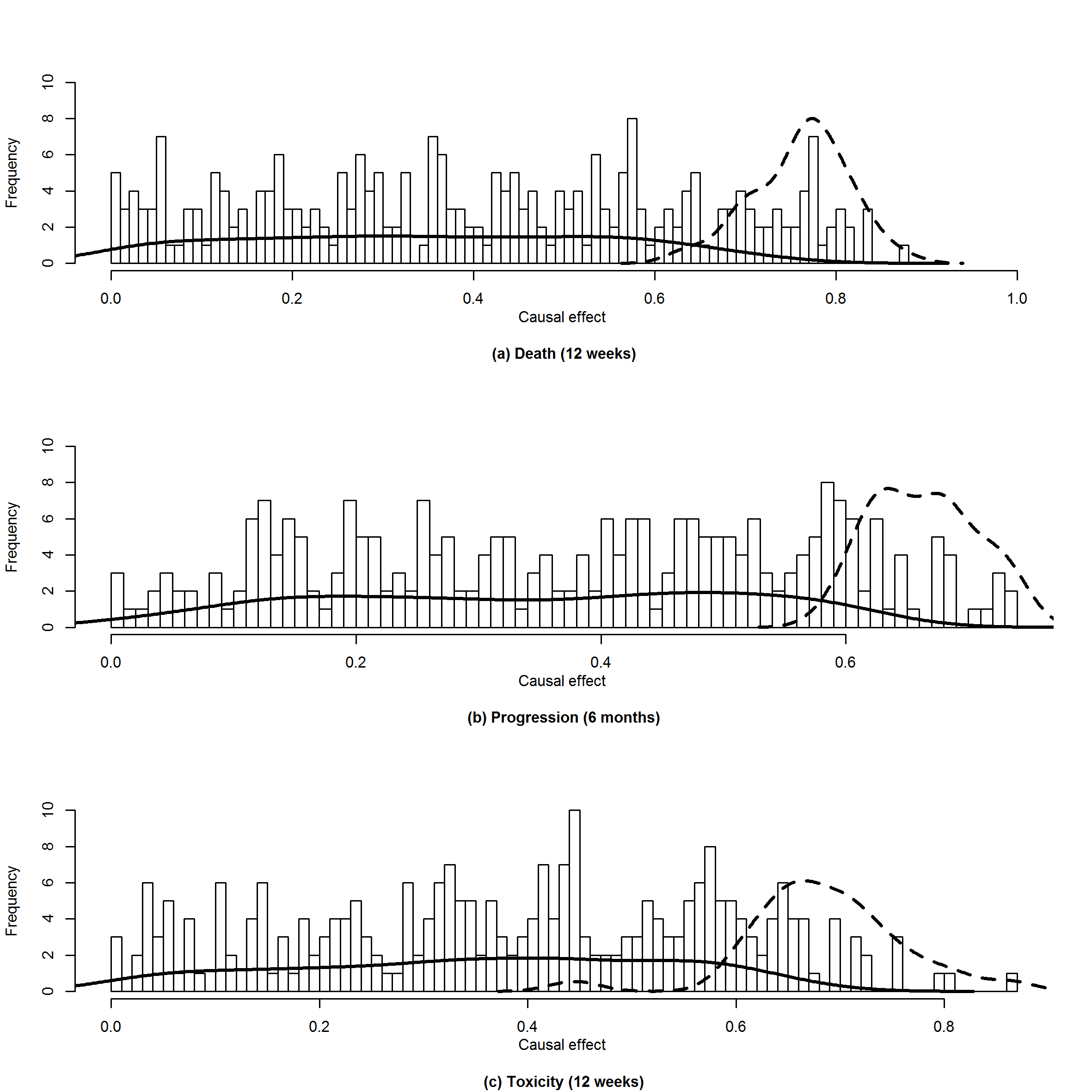}
        \caption{Histogram of the causal effect for the biomarkers on death, progression and toxicity based on model 2 over 300 runs. Solid and dashed lines represent the kernel density of biomarkers with a PCER $> 0.5\%$ and PCER $\leq 0.5\%$ respectively. \label{Model2} }
         
\end{figure}

\begin{figure}[h!]
\centering   
        \includegraphics[scale=0.65]{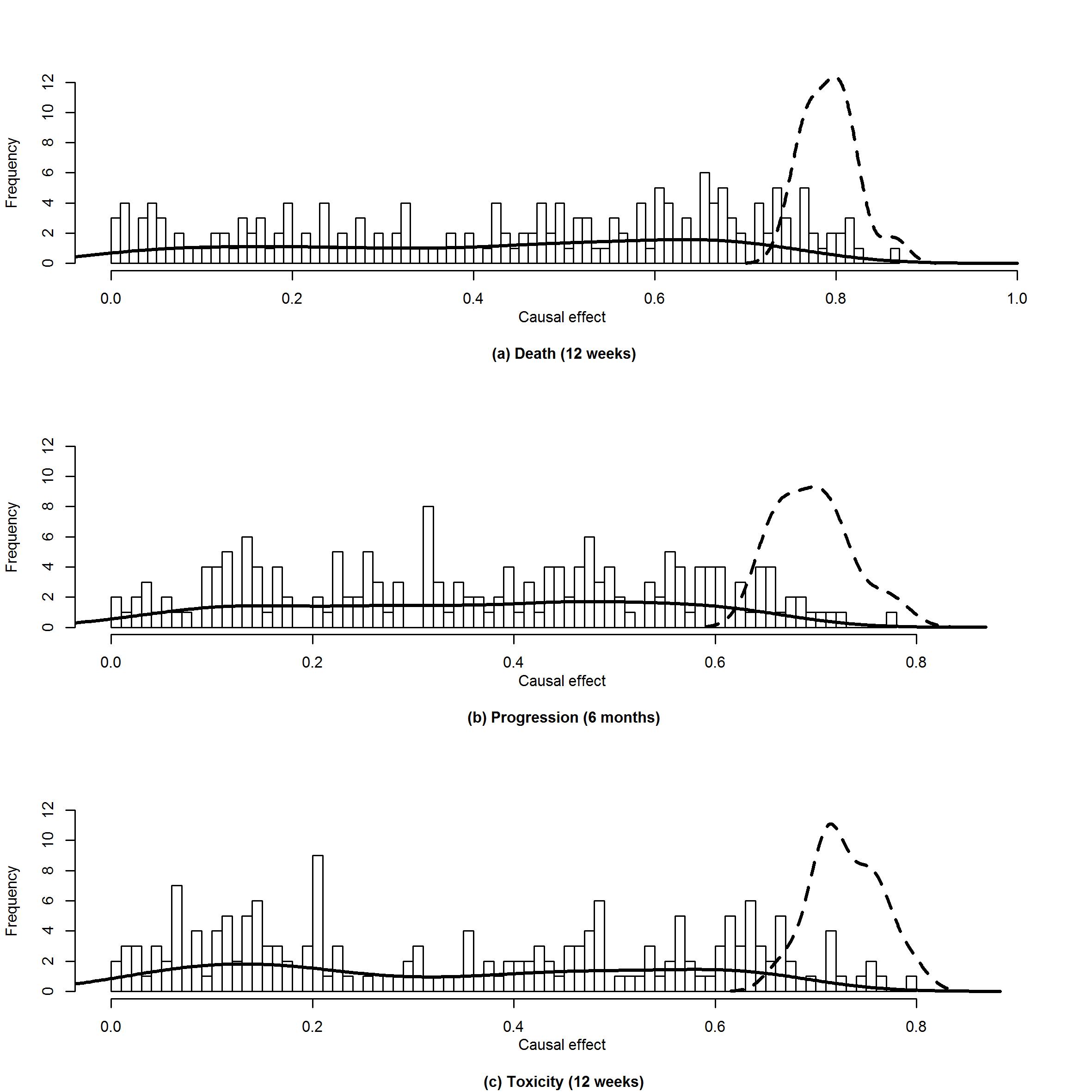}
        \caption{Histogram of the causal effect for the biomarkers on death, progression and toxicity based on model 3 over 300 runs. Solid and dashed lines represent the kernel density of biomarkers with a PCER $> 0.5\%$ and PCER $\leq 0.5\%$ respectively.\label{Model3}}

\end{figure}

\begin{table}[h!]
\centering
\caption{Top 10 of immunological biomarkers with a PCER $< 0.5\%$ in model 2. The number following “v” stands for the visit number. Superscript indicate biomarkers in common. See appendix C for the complete description of the biomarkers.\label{my-label}}

\scalebox{0.85}{
\begin{tabular}{cccccccccc}
\hline
\multirow{2}{*}{Rank} & \multicolumn{3}{c}{Death (12 weeks)} & \multicolumn{3}{c}{Progression (6 months)} & \multicolumn{3}{c}{Toxicity 12 weeks} \\ \cline{2-10} 
 & Biomarker & \begin{tabular}[c]{@{}c@{}}Median \\ effect\end{tabular} & PCER & Biomarker & \begin{tabular}[c]{@{}c@{}}Median \\ effect\end{tabular} & PCER & Biomarker & \begin{tabular}[c]{@{}c@{}}Median\\  effect\end{tabular} & PCER \\ \hline
1 & BM16v2$^{a}$ & 0.81 & 0.0035 & BM8v1$^{d}$ & 0.77 & 0.0031 & BM7v4 & 0.79 & 0.0028 \\ 
2 & BM5v1 & 0.81 & 0.0035 & BM44v4 & 0.72 & 0.0036 & BM8v4$^{d}$ & 0.76 & 0.0034 \\ 
3 & BM42v3 & 0.8 & 0.0037 & BM26v2 & 0.71 & 0.0041 & BM16v3$^{a}$ & 0.75 & 0.0036 \\ 
4 & BM48v1 & 0.86 & 0.0037 & BM30v3$^{b}$ & 0.71 & 0.0041 & BM26v4 & 0.75 & 0.0036 \\ 
5 & BM42v2 & 0.8 & 0.0038 & BM44v3 & 0.68 & 0.0042 & BM7v3 & 0.76 & 0.0037 \\ \
6 & BM14v4 & 0.79 & 0.0039 & BM45v1$^{e}$ & 0.7 & 0.0047 & BM9v4$^{c}$ & 0.72 & 0.0039 \\ 
7 & BM30v4$^{b}$ & 0.8 & 0.0039 & BM39v4$^{f}$ & 0.66 & 0.0049 & BM39v3$^{f}$ & 0.72 & 0.0039 \\ 
8 & BM11v4 & 0.83 & 0.004 & BM40v4 & 0.66 & 0.0049 & BM32v3 & 0.71 & 0.0042 \\ 
9 & BM11v1 & 0.76 & 0.0042 & BM14v2 & 0.66 & 0.005 & BM30v1$^{b}$ & 0.67 & 0.0045 \\ 
10 & BM9v4$^{c}$ & 0.81 & 0.0043 & - & - & - & BM45v1$^{e}$ & 0.71 & 0.0046 \\ \hline
\end{tabular}
}
\end{table}
We see that some of the biomarkers are present in all top 10 but differ with the time of measurement. We see that BM30 is present in the top 10 of toxicity at visit 1, in the top 10 of progression at visit 3 and in the top ten of the death at visit 4.  Other biomarkers are present in 2 of the top 3 but differ with the visit such as BM26, BM45, BM39 and BM9.
\newpage

\section{Discussion}
We extended in this paper the IDA method to repeated measures by introducing a chronologically ordered (CO) version of the so called PC-algorithm. Our proposed algorithm COPC-algorithm takes a priori chronological information such as repeated measure into account in the input graph. We applied then PC-stable and our new method COPC-stable to simulated data sets and observational data of repeated immunological biomarkers from patients treated repeatedly with immunotherapy for metastatic melanoma. When comparing CPDAGs obtained with PC-stable and those with COPC-stable, the simulation study showed that PC-stable had a lower sensitivity than the COPC-stable leading to a better learning of the true structure. On the application, CPDAGs based on PC-stable had indeed non-chronological ordered paths while those based on COPC-stable could not have any. CPDAGs obtained with COPC-stable had on average more total and directed edges than those obtained with PC-stable but less bidirected edges. The lower the number of directed edges, the lower the number of possible ways to direct edges, and hence the lower the number of DAGs in the \textit{Markov equivalence class}. Moreover Table 4 showed that when using COPC-stable, the proportion of values obtained in the multiset $\theta_{i}$  was on average lower when using PC-stable. Smaller is the \textit{Markov equivalence class}, higher is the power of the study to identify causal effects.\\ 
In the COPC-stable, the number of tested conditional dependencies is considerably smaller than with PC-stable. Since it takes chronological order information into account, the COPC-algorithm does not test dependencies of two variables conditioning on a variable measured at a time after those two variables. In contrary, the original PC-algorithm tests non-realistic conditional dependences and thus raises the number of global tests. Testing those non-realistic conditional dependences could lead to identify false positive causal effects.\\ 
Finding the true causal DAG has always been the principle interest of causal inference studies, knowing the true causal DAG allows estimating the true causal effect. However, in high-dimensional setting, the true causal DAG is generally unknown and it is difficult to check whether or not all possible confounders are measured. In this case, causal effects cannot be uniquely estimated. Therefore IDA was developed to estimate lower bounds of the causal effects of $X_{i}$ on Y and determine the importance of these effects.  This is a different approach where instead of searching one true causal effect, a range of  causal effects are estimated in each DAG from a \textit{Markov equivalence class}. Consequently, when effect of large numbers of markers is identified, those which have causal effects could be selected by different approaches. In fact, we could either keep a small range of biomarkers that are in the top effects as in \cite{Stekhoven2012} or a larger range of those with a limited but slightly higher probability of being false positive. In the high-dimensional setting, the first approach will keep biomarkers with the strongest causal effect but not necessarily all biomarkers with a small causal effect. The second approach assures to select a larger list of biomarkers that have a robust causal effect and will suggest to clinicians which immunological biomarkers they should investigate deeper in a follow-up study. Also, controlling for type 1 error can be done by different methods. We choose in our application the PCER because it is less restrictive compared to methods such as FDR (False discovery rate) or FWER (Family-wise error rate).\\
The choice of the selecting approach depends on the objective: selecting a small list of biomarker that have the highest effect on the outcome or identifying all the biomarkers that have an effect regardless of the size effect. For instance, if only the measure of a marker at visit 2 belongs to the top causal effects, is only the measure at visit 2 is important or should the marker be measured at all visits ?\\ Usually, in a causal DAG, all true causal effects have to be reported, not only the strongest. Nevertheless, the interpretation of the top causal biomarker is challenging. Having a biomarker at a certain visit with a PCER below the selected threshold does not mean that the biomarker has a causal effect only at this visit but rather its maximum and more robust effect at this visit. 
One of the main assumptions made in this study is that the true DAG is not dynamic like other extensions of the PC-algorithm on time-series data \cite{Chu2008,Brodersen2015a}. So we did not constrain the arrows to be the same within each visit. In fact, the context of biological biomarkers can be much more complex than a simple repetition of a pattern. Originally, the IDA made the assumption that all variables including the outcome were Gaussian, then it has been extended in a case where all variables (including outcome) are discrete \cite{Kalisch2010a}. In this study we made the assumption that all covariates $X=\{X_{1},…,X_{p} \}$ are Gaussian and that the outcome is binary because it is a situation that is quite common in oncology. Also, the covariates need to be measured at uniform set of time points (i.e. balanced data). \\   
Our work was motivated by finding causal effects among repeated immunological biomarkers on death and toxicity of patients treated with immunotherapy for metastatic melanoma. Based on our observational data, using the IDA with our new version of the PC-algorithm, the COPC-algorithm, we found a consistent list of immunological biomarkers with causal effects. But one should be attentive not to overinterpret these results. It is in fact impossible in an accurate way to check whether or not our assumptions hold; having no unmeasured confounders is a strong assumption but may be reasonable in our application.  \\
Further work will investigate the adding of expert knowledge as input of the COPC-algorithm based on high-dimensional graphs. Also we will explore extensions that can deal with longitudinal and time to event outcomes.

\section{Conclusion}
We presented in this paper, an extension of the PC-algorithm called COPC-algorithm. It provides CPDAGs that keep the chronological structure present in the data and allow us thus to estimate reliable lower bounds of the causal effect of repeated covariates or biomarkers. In the immunotherapy example, immunological biomarkers on early toxicity, premature death and progression were identified and will be further investigated by clinicians.

\section{Declarations}
\subsection*{Ethics approval  and consent to participate}
Patients were informed of the study and consented to participate. This study was approved by the Kremlin Bicêtre Hospital Ethics Committee (SC12-018; ID-RCB-2012-A01496-37) and all procedures were performed in accordance with the Declaration of Helsinki.
\subsection*{Competing interest}
The authors declare that they have no competing interests.
\subsection*{Funding}
This work is part of a PhD thesis funded by Université Paris Sud.\\
This study was funded by Gustave Roussy Cancer Campus, Fondation Gustave Roussy, the Institut national de la santé et de la recherche médicale (INSERM), the Direction Générale de l’Offre de Soins  (DGOS; GOLD TRANSLA 12-174), the Institut National du Cancer (INCa; GOLD 2012-062 N Cancéropôle : 2012-1-RT-14-IGR-01), and SIRIC SOCRATE (INCa DGOS INSERM 6043), MMO program: ANR-10IBHU-0001).
\subsection*{Authors' contributions}
VA, SM, EM have conceived the statistical work, VA has drafted the manuscript. CC and NC provided immunological data and helped for the immunological interpretation. CR provided clinical data and help for clinical interpretation. All authors have critically reviewed the manuscript. 
\subsection*{Acknowledgements}
Vahé Asvatourian was supported by Université Paris Sud. Clelia  Coutzac  was  supported  by  fellowships  from  Fondation  pour  la  Recherche  Médicale (FRM).The authors would like to thank Gustave Roussy Cancer Campus, INSERM and INCa for their funding.

 \newpage
 \begin{appendices}
  

\section*{Supplementary material}

# Appendix A Estimation of the causal effect of a repeated biomarker on a binary outcome

Causal effects of $X_i$ on the outcome $Y$ can be quantified by measuring the difference of any function of the distribution of "counterfactuals outcome" such as mean, median or ratios (see technical point 1.1 of [1]). Let $G$ be a directed acyclic graph with V the set of nodes described by $p + 1$ variables $X_1, \ldots, X_p, Y$.

Pearl [2] showed that the distribution generated from a DAG is called Markovian and can be factorized as

(1) $$P(x_1, \ldots, x_p, Y = x_p + 1) = \prod_{i=1}^{p+1} P(x_i | pa(x_i)).$$

Intervention on a variable using Pearl's do operator assigns a value to the variable over the population; and the distribution generated on the variables set can be expressed in the *truncated factorization formula*:

(2) $$P(x_1, \ldots, x_p, Y = x_p + 1 | do(X_j = x'_j)) = \begin{cases} \prod_{i=1, i \neq j}^{p+1} P(x_i | pa(x_i)) & \text{if } x_j = x'_j \\ 0 & \text{if } x_j \neq x'_j. \end{cases}$$

Equation (2) reflects the removal in (3) of $P(x_i|pa(x_i))$ since $pa(x_i)$ has no effect on $x_i$. Graphically, removing $P(x_i|pa(x_i))$ is equivalent to removing the arrows from $pa(x_i)$ to $x_i$. Based on (2), Pearl's has shown that the effect of the intervention $do(X_i = x'_i)$ on Y is given by

(3) $$P(Y|do(X_j = x'_i)) = \sum_{pa(x_i)} P(Y|do(X_i = x'_i), pa(x_i)) \, P(pa(x_i)).$$

Equation (3) means conditioning $P(Y|do(X_j = x'_i))$ on the parents of $X_i$ and then averaging the results weighted by the probability of $pa(x_i)$.

For a continuous outcome, $P(Y|do(X_j = x'_i)) = E(Y|do(X_i = x))$, which denotes the mean of $Y$ when $X_i$ is uniformly assigned to $x$ over the population. So we can defined the average causal effect for a continuous outcome $Y$ by

(4) $$E(Y|do(X_i = x)) - E(Y|do(X_i = x'_i)),$$

where $E(Y|do(X_i = x_i))$ and $E(Y|do(X_i = x'_i))$ denote the mean of $Y$ when $X_i$ is uniformly assigned to $x_i$ or $x'_i$ over the population via Pearl's *do* operator. In the case of continuous Gaussian variables, the causal effect of $X_i$ on $Y$ is the regression coefficient $\beta_1$ of $X_i$ in the linear regression of $Y$ on $X_i$ and $pa(X_i, G)$ [1, 2]:

(5) $$E(Y|X_i, pa(X_i, G)) = \beta_0 + \beta_1 X_i + \beta_{pa_i} pa_i.$$

However, in the case where covariates $X_1, \ldots, X_p$ are Gaussian and the outcome Y is binary, the linearity stated in (5) does not hold. This is why we modelled the relation using a logit link that allows keeping the linearity and calculating the causal effect for a binary outcome as

(4) $$logit\, P(Y = 1|X_i, pa(X_i, G)) = \beta_0 + \beta_1 X_i + \beta_{pa_i} pa_i.$$

This method of modelling is widely used in [1] (see technical point 11.1).

However, the "separation" phenomenon [3] could occur in small datasets in logistic regression such as in our melanoma example: subjects having $y = 1$ and subjects having $y = 0$ can be separated by a single or a combination of covariates, the likelihood converges while at least one parameter estimate diverges to $\pm\infty$, leading to infinite odds ratio estimates. To overcome this situation, Firth proposed to reduce the bias of maximum likelihood estimates [4]. Several studies have shown that this method provides unbiased estimates [5, 6]. In our small observational dataset setting, the causal effects on dichotomous outcome will be estimated through logistic regression with Firth correction.

# Appendix B Simulation set-up

In this additional file, we explain how we run our simulations and what scenarios we tested. In our scenarios, we fixed the number of variables, of maximum parent per node and the correlation between biomarkers. We tested different value for the number of observations, the number of measurement (visits) and for the independence test cut-off alpha leading to a total of 8 scenarios.

## Scenarios

We generated data faithful to a by using the following parametrisations:

- $p = 20$
- $n_{visits} \in \{4, 6\}$
- $maxP = 3$
- $n \in \{50, 1000\}$
- $\alpha \in \{0.02, 0.2\}$
- $\rho_{min} = 0.5$, $\rho_{max} = 0.7$
- $\sigma = 1$

For each scenario we generated a DAG with $p * n_{vis} + Y$ variables and then sampled 500 different datasets per DAG. We calculated the average SHD, sensibility and specificity for estimated CPDAGs with both CP-stable and COPC-stable based on the 500 datasets.

## Generate a DAG with repeated measures of covariates

To generate a DAG that has repeated measures and an expected number of parents per node $maxP$, we used the following approach: first, we generated a $(p \times n_{visits} + Y) \times (p \times n_{visits} + Y)$ weighted matrix $wM$, where $p$ is the number of biomarkers, $n_{visits}$ the number of visits (measurements) and $Y$ the outcome; with independent realizations of $Uniform(0.5, 1)$ in the upper triangle of the matrix and zeroes in the remaining entries. At the end, if $wM_{i,j} > 0$ then, it meant there was an arrow from $X_i$ to $X_j$ ($X_i \to X_j$) with a weight of $wM_{i,j}$. The maximum parent for each node was limited by $maxP$. Then, based on the true DAG obtained, we generated i.i.d samples using algorithm 1.

**Algorithm 1:** Generation of data based on a DAG with repeated measures of covariates with a single outcome

**Input:** number of observations ($n$), weighted matrix ($wM$), sigma ($\sigma$), $\rho_{min}$, $\rho_{max}$, number of biomarkers ($p$), number of measurements ($n_{visits}$)

**for** Subject $i \leftarrow 1$ **to** n **do**

    **for** Number of Biomarkers $j \leftarrow 1$ **to** $p$ **do**

        Let $V_j$ be the set of measurements for the biomarker $j$

        Let the vector $\mu_{V_j} = E(V_1) = E(V_2) = E(V_{n_{visits}}) = 0$

        $\rho_j \sim Uniform(\rho_{min}, \rho_{max})$

$$\Sigma_j = \begin{pmatrix} \sigma^2 & \rho^1\sigma^2 & \rho^2\sigma^2 & \cdots & \rho^{n_{visits}}\sigma^2 \\ \rho^1\sigma^2 & \sigma^2 & \rho^1\sigma^2 & \cdots & \rho^{n_{visits}-1}\sigma^2 \\ \rho^2\sigma^2 & \rho^1\sigma^2 & \sigma^2 & \cdots & \rho^{n_{visits}-2}\sigma^2 \\ \vdots & \vdots & \vdots & \ddots & \vdots \\ \rho^{n_{visits}}\sigma^2 & \rho^{n_{visits}-1}\sigma^2 & \rho^{n_{visits}-2}\sigma^2 & \cdots & \sigma^2 \end{pmatrix} \quad \epsilon_{i,V_j} \sim N_{n_{visits}}(\mu_j, \Sigma_j)$$

        $\epsilon_{i,Outcome} \sim N(0,1)$

    **end**

**end**

$p_{tot} = p \times n_{visits} + 1$ (outcome)

**for** Subject $i \leftarrow 1$ **to** n **do**

    **for** Variable $v \leftarrow 1$ **to** $p_{tot}$ **do**

        **if** $Variable_v$ has no parent **then**

            $Variable_{i,v} = \epsilon_{Variable_{i,v}}$

        **end**

        **else if** $Variable_v$ has at least 1 parent **then**

            $Variable_{i,v} = \sum_{j=1}^{n_{parents}}(wM_{j,v} \times PA_j(Variable_{i,v})) + \epsilon_{Variable_{i,v}}$

            **if** $v = p$ **then**

                $Variable_{i,p} \sim Bernoulli(logit^{-1}(Variable_{i,p_{tot}}))$

            **end**

        **end**

    **end**

**end**

**Output:** Simulated database

# Appendix C Description of all anonymised biomarkers of the application

**Table 1:** Description of the anonymised biomarkers for model 1. The number following "v" stands for the visit number.

| Labels | Description |
|---|---|
| **BM 1v1** | Interleukin 6 |
| **BM 2 v1** | Interleukin 8 |
| **BM 3 v1** | Monocyte chemo attractant protein 1 |
| **BM 4 v1** | Interferon gamma-induced protein 10 |
| **BM 5 v1** | Tumour Necrosis Factor $\alpha$ |
| **BM 6 v1-v4** | C reactive protein |
| **BM 7 v1-v4** | Lactate Dehydrogenase |
| **BM 8 v1-v4** | Eosinophils |
| **BM 9 v1-v4** | Neutrophils |
| **BM 10 v1-v4** | Soluble CD25 |
| **BM 11 v1-v4** | Soluble CTLA-4 |
| **BM 12 v1-v4** | $CD4^+$ T cells |
| **BM 13 v1-v4** | $CD8^+$ T cells |

**Table 2:** Description of the anonymised biomarkers for model 3. The number following "v" stands for the visit number.

| Labels | Description |
|---|---|
| **BM 1v1** | Interleukin 6 |
| **BM 2 v1** | Interleukin 8 |
| **BM 3 v1** | Monocyte chemo attractant protein 1 |
| **BM 4 v1** | Interferon gamma-induced protein 10 |
| **BM 5 v1** | Tumour Necrosis Factor $\alpha$ |
| **BM 6 v1-v4** | C reactive protein |
| **BM 7 v1-v4** | Lactate Dehydrogenase |
| **BM 8 v1-v4** | Eosinophils |
| **BM 9 v1-v4** | Neutrophils |
| **BM 10 v1-v4** | Soluble CD25 |
| **BM 11 v1-v4** | Soluble CTLA-4 |
| **BM 14 v1-v4** | % of $Foxp3^+CD25^+$ among $CD4^+$ T cells |
| **BM 15 v1-v4** | % of conventional $CD4^+$ T cells (Tconv) ($FoxP3^-$) |
| **BM 16 v1-v4** | % of $CD49d^+\beta7^+$ among $CD4^+$ T cells |
| **BM 17 v1-v4** | % of $CD49d^+\beta7^+$ among $CD4^+$ T cells |
| **BM 18 v1-v4** | % of $\beta7^+CD103^+$ among $CD4^+$ T cells |
| **BM 19 v1-v4** | % of C-C chemokine receptor $6^+(CCR6^+)$ among $CD4^+$ T cells |
| **BM 20 v1-v4** | % of Cutaneous Lymphocyte-associated antigen$^+$ ($CLA^+$) C-C chemokine receptor |

| | | |
|---|---|---|
| | | $10^+$ (CCR10$^+$) among CD4$^+$ T cells |
| **BM 21** | **v1-v4** | % of CLA$^+$ among CD4$^+$ T cells |
| **BM 22** | **v1-v4** | % of central memory among CD4$^+$ T cells |
| **BM 23** | **v1-v4** | % of C-X-C chemokine receptor 3$^+$(CXCR3$^+$) CCR6$^+$ among CD4$^+$ T cells |
| **BM 24** | **v1-v4** | % of CXCR3$^+$ chemokine receptor+ among CD4$^+$ T cells |
| **BM 25** | **v1-v4** | % of C-X-C chemokine receptor 5$^+$ (CXCR5$^+$) chemokine receptor$^+$ among CD4$^+$ T cells |
| **BM 26** | **v1-v4** | % of effector memory among CD4$^+$ T cells |
| **BM 27** | **v1-v4** | % of effector among CD4$^+$ T cells |
| **BM 28** | **v1-v4** | % of memory among CD4$^+$ T cells |
| **BM 29** | **v1-v4** | % of naïve among CD4$^+$ T cells |
| **BM 30** | **v1-v4** | % of CD49d$^+$ CD103$^+$ among CD8$^+$ T cells |
| **BM 31** | **v1-v4** | % of CD49d$^+$ β7$^+$ among CD8$^+$ T cells |
| **BM 32** | **v1-v4** | % of β7$^+$CD103$^+$ among CD8$^+$ T cells |
| **BM 33** | **v1-v4** | % of CCR10$^+$ among CD8$^+$ T cells |
| **BM 34** | **v1-v4** | % of CCR6$^+$ among CD8$^+$ T cells |
| **BM 35** | **v1-v4** | % of CLA$^+$ CCR10$^+$ among CD8$^+$ T cells |
| **BM 36** | **v1-v4** | % of CLA$^+$ among CD8$^+$ T cells |
| **BM 37** | **v1-v4** | % of central memory among CD8 T cells |
| **BM 38** | **v1-v4** | % of CXCR3$^+$ CCR6$^+$ among CD8$^+$ T cells |
| **BM 39** | **v1-v4** | % of CXCR3$^+$ among CD8$^+$ T cells |
| **BM 40** | **v1-v4** | % of CXCR5$^+$ among CD8$^+$ T cells |
| **BM 41** | **v1-v4** | % of effector memory among CD8$^+$ T cells |
| **BM 42** | **v1-v4** | % of effector among CD8$^+$ T cells |
| **BM 43** | **v1-v4** | % of memory among CD8$^+$ T cells |
| **BM 44** | **v1-v4** | % of naïve among CD8$^+$ T cells |
| **BM 45** | **v1-v4** | % of Inducible CoStimulator (ICOS$^+$) among CD8$^+$ T cells |
| **BM 46** | **v1-v4** | % of Inducible CoStimulator (ICOS$^+$) among CD4$^+$ T cells |
| **BM 47** | **v1-v4** | % of Inducible CoStimulator Ligand (ICOSL$^+$) among CD4$^+$ T cells |
| **BM 48** | **v1-v4** | % of High Inducible CoStimulator Ligand (ICOSL$^+$) among CD4$^+$ T cells |
| **BM 49** | **v1-v4** | % of Inducible CoStimulator Ligand (ICOSL$^+$) among CD8$^+$ T cells |
| **BM 50** | **v1-v4** | % of FoxP3$^+$ among CD4$^+$ T cells |

**Table 3:** Description of the anonymised biomarkers for model 3. The number following "v" stands for the visit number.

| Labels | Description |
| --- | --- |
| **BM 1v1** | Interleukin 6 |
| **BM 2 v1** | Interleukin 8 |
| **BM 3 v1** | Monocyte chemo attractant protein 1 |
| **BM 4 v1** | Interferon gamma-induced protein 10 |
| **BM 5 v1** | Tumour Necrosis Factor $\alpha$ |
| **BM 6 v1-v4** | C reactive protein |
| **BM 7 v1-v4** | Lactate Dehydrogenase |
| **BM 8 v1-v4** | Eosinophils |
| **BM 9 v1-v4** | Neutrophils |
| **BM 10 v1-v4** | Soluble CD25 |
| **BM 11 v1-v4** | Soluble CTLA-4 |
| **BM 51 v1-v4** | % of $CD49d^+CD103^+$ among central memory $CD4^+$ T cells |
| **BM 52 v1-v4** | % of $CD49d^+\beta7^+$ among central memory $CD4^+$ T cells |
| **BM 53 v1-v4** | % of $\beta7^+CD103^+$ among central memory $CD4^+$ T cells |
| **BM 54 v1-v4** | % of $CD49d^+CD103^+$ among effector memory $CD4^+$ T cells |
| **BM 55 v1-v4** | % of $CD49d^+\beta7^+$ among effector memory $CD4^+$ T cells |
| **BM 56 v1-v4** | % of $\beta7^+CD103^+$ among effector memory $CD4^+$ T cells |
| **BM 57 v1-v4** | % of $CD49d^+CD103^+$ among effector $CD4^+$ T cells |
| **BM 58 v1-v4** | % of $CD49d^+\beta7^+$ among effector $CD4^+$ T cells |
| **BM 59 v1-v4** | % of $\beta7^+CD103^+$ among effector $CD4^+$ T cells |
| **BM 60 v1-v4** | % of $CCR10^+$ among effector $CD4^+$ T cells |
| **BM 61 v1-v4** | % of $CCR\ 6^+$ among effector $CD4^+$ T cells |
| **BM 62 v1-v4** | % of $CLA^+CCR10^+$ among effector $CD4^+$ T cells |
| **BM 63 v1-v4** | % of $CLA^+$ among effector $CD4^+$ T cells |
| **BM 64 v1-v4** | % of $CXCR3^+\ CCR6^+$ among effector $CD4^+$ T cells |
| **BM 65 v1-v4** | % of $CXCR3^+$ among effector $CD4^+$ T cells |
| **BM 66 v1-v4** | % of $CXCR5^+$ among effector $CD4^+$ T cells |
| **BM 67 v1-v4** | % of CD49+b7+ among memory $CD4^+$ T cells |
| **BM 68 v1-v4** | % of $\beta7^+CD103^+$ among memory $CD4^+$ T cells |
| **BM 69 v1-v4** | % of $CCR10^+$ among memory $CD4^+$ T cells |
| **BM 70 v1-v4** | % of $CCR6^+$ among memory $CD4^+$ T cells |
| **BM 71 v1-v4** | % of $CLA^+\ CCR10^+$ among memory $CD4^+$ T cells |
| **BM 72 v1-v4** | % of $CLA^+$ among memory $CD4^+$ T cells |
| **BM 73 v1-v4** | % of $CXCR3^+\ CCR6^+$ among memory $CD4^+$ T cells |
| **BM 74 v1-v4** | % of $CXCR3^+$ among memory $CD4^+$ T cells |
| **BM 75 v1-v4** | % of $CXCR5^+$ among memory $CD4^+$ T cells |
| **BM 76 v1-v4** | % of $CD49d^+CD103^+$ among naïve $CD4^+$ T cells |
| **BM 77 v1-v4** | % of $CD49d^+\beta7^+$ among naïve $CD4^+$ T cells |
| **BM 78 v1-v4** | % of $\beta7^+CD103^+$ among naïve $CD4^+$ T cells |
| **BM 79 v1-v4** | % of $CD49d^+CD103^+$ among central memory $CD8^+$ T cells |
| **BM 80 v1-v4** | % of $CD49d^+\beta7^+$ among central memory $CD8^+$ T cells |
| **BM 81 v1-v4** | % of $\beta7^+CD103^+$ among central memory $CD8^+$ T cells |
| **BM 82 v1-v4** | % of $CD49d^+CD103^+$ among effector memory $CD8^+$ T cells |

| | | |
|---|---|---|
| **BM 83 v1-v4** | | % of CD49d$^+$β7$^+$ among effector memory CD8$^+$ T cells |
| **BM 84 v1-v4** | | % of β7$^+$CD103$^+$ among effector memory CD8$^+$ T cells |
| **BM 85 v1-v4** | | % of CD49d$^+$CD103$^+$ among effector CD8$^+$ T cells |
| **BM 86 v1-v4** | | % of CD49d$^+$β7$^+$ among effector CD8$^+$ T cells |
| **BM 87 v1-v4** | | % of β7$^+$CD103$^+$ among effector CD8$^+$ T cells |
| **BM 88 v1-v4** | | % of CCR10$^+$ among effector CD8$^+$ T cells |
| **BM 89 v1-v4** | | % of CCR6$^+$among effector CD8$^+$ T cells |
| **BM 90 v1-v4** | | % of CLA$^+$ CCR10$^+$ among effector CD8$^+$ T cells |
| **BM 91 v1-v4** | | % of CLA$^+$ among effector CD8$^+$ T cells |
| **BM 92 v1-v4** | | % of CXCR3$^+$ CCR6$^+$ among effector CD8$^+$ T cells |
| **BM 93 v1-v4** | | % of CXCR3$^+$ among effector CD8$^+$ T cells |
| **BM 94 v1-v4** | | % of CXCR5$^+$among effector CD8$^+$ T cells |
| **BM 95 v1-v4** | | % of CCR10$^+$ among memory CD8$^+$ T cells |
| **BM 96 v1-v4** | | % of CCR6$^+$ among memory CD8$^+$ T cells |
| **BM 97 v1-v4** | | % of CLA$^+$ CCR10$^+$ among memory CD8$^+$ T cells |
| **BM 98 v1-v4** | | % of CLA$^+$ among memory CD8$^+$ T cells |
| **BM 99 v1-v4** | | % of CXCR3$^+$ CCR6$^+$ among memory CD8$^+$ T cells |
| **BM 100 v1-v4** | | % of CXCR3$^+$ among memory CD8$^+$ T cells |
| **BM 101 v1-v4** | | % of CXCR5$^+$ among memory CD8$^+$ T cells |
| **BM 102 v1-v4** | | % of CD49d$^+$CD103$^+$ among naïve CD8$^+$ T cells |
| **BM 103 v1-v4** | | % of CD49d$^+$β7$^+$ among naïve CD8$^+$ T cells |
| **BM 104 v1-v4** | | % of β7$^+$CD103$^+$ among naïve CD8$^+$ T cells |
| **BM 105 v1-v4** | | % of ICOS$^+$ cells among Tconv cells |
| **BM 106 v1-v4** | | % of ICOSL$^+$ cells among Tconv cells |
| **BM 107 v1-v4** | | % of ICOSL$^+$ cells among Treg cells |
| **BM 108 v1-v4** | | % of ICOS$^+$ cells among Treg cells |

# Appendix D The use of missingness graph for repeated measurements of multi-dimensional biomarkers

In studies with repeated biomarker data, missing data could be either MCAR (missing completely at random), MAR (missing at random) or MNAR (missing not at random) [1, 2]. To represent the causal mechanisms underlying in each category and the corresponding assumption about their causal impact, Mohan et al proposed the use of graphical model using conditional independencies[3, 4]. The graphical models used in this case are called missingness graphs (*m-graphs*). These graphs are an efficient way of presenting the properties of the missingness mechanisms and thus, the potential of recovering missing data. Let $G(V, E)$ be the DAG where $V$ is the set of observable nodes and $E$ the set of edges in the DAG. V can be separated into $V_{obs}$ and $V_{mis}$ where $V_{obs}$ is the set of variables that are fully observed and $V_{mis}$ is the set of variables that are missing in at least one record. Let $V_i$ a variable of interest and $V_i^*$ the variable which is actually observed, $R_{v_i}$ is the causal missingness mechanism of $V_i^*$. They also introduce the notion of recoverability [3] where under some conditions, an un biased estimate of given relation $Q$ can be computed. If data D are generated by a process compatible with a graph G, a procedure that computes an estimator $\hat{Q}(D)$ of the relation $Q$ converges to $Q$ in the limit of large samples.

The reason why it is important to determine the missingness mechanism of a variable is, because missing data due to MCAR, MAR or MNAR need different approaches. For instance, MCAR data can be listwise deleted or simple imputed. For MAR, multiple imputation can provide consistent estimates while pattern mixture models seem to be most appropriate when data are MNAR [2].

Missingness mechanisms and their recoverability can be expressed in the following way:

- Missing completely at random (MCAR)

Data are called MCAR when the probability that $V_i$ is missing is independent from all other variables: $P(R_{v_i}|V_{obs}, V_{mis}) = P(R_{v_i}) \Leftrightarrow R_{v_i} \perp (V_{obs}, V_{mis})$. Thus $P(V) = P(V|R) = P(V_{obs}, V^*|R = 0)$. Since R and $V^*$ are currently observed, the joint distribution $P(V)$ is recoverable. Figure 1 shows an example of a MCAR model where $A$ is an auxiliary variable fully observed, and $X$ the variable with missing values. In this example, based on d-separation notion, the missingness mechanism $R_x$ is independent of all missing and fully observed variables such as X and A: $R_x \perp (A, X)$. The joint distribution $P(X, A)$ is then recoverable.

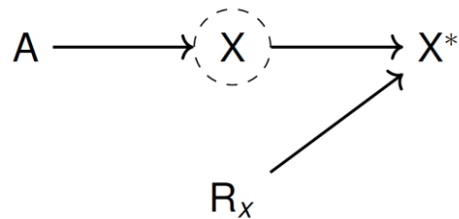

**Figure 1:** A simple MCAR model. Nodes with dashed circle represent variables that would have been observed had they not missing values. Nodes with a star represent observed variables with missing values.

- Missing at random (MAR)

Data are called MAR when the missingness mechanism of $V_i$ is conditionally independent of $V_i$: $P(R_{v_i}|V_{obs}, V_{mis}) = P(R_{v_i}|V_{obs}) \Leftrightarrow R_{v_i} \perp V_{mis}|V_{obs}$, thus $P(V) = P(V_{mis}|V_{obs})P(V_{obs}) = P(V^*|V_{obs}, R = 0)P(V_{obs})$. Since R and $V^*$ are currently observed, the joint distribution $P(V)$ is recoverable. Figure 2 shows an example of a simple MAR model. In this example, based on the d-separation notion, the missingness mechanism $R_x$ and $X$ are d-connected via $A$. But, as long as $A$ is fully observed and if conditioning on $A$ blocks the path between $X$ and $R_x$, then $R_x$ is conditionally independent of $X$ knowing $A$ ($R_x \perp X|A$). The joint distribution $P(X, A)$ is then recoverable. In this case, conditioning on a variable refers to use this variable as a predictor in a multiple imputation model [4].

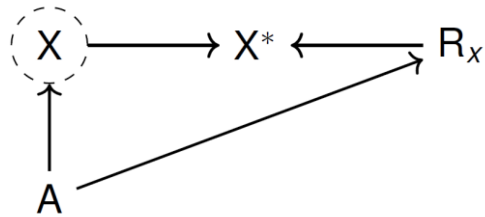

**Figure 2:** A simple MAR model. Nodes with dashed circle represent variables that would have been observed had they not missing values. Nodes with a star represent observed variables with missing values.

- Missing not at random (MNAR)

Data are MNAR when neither MCAR nor MAR. This can occur when the probability of a missingness mechanism is dependant of another variable: $P(R_{v_i}|V_{obs}, V_{mis}) \neq P(R_{v_i}|V_{obs})$. Figure 3 shows some typical situations where data are MNAR. Figure 3a shows the classical situation of MNAR data when there is a direct path between the missing variable and its missingness mechanism. Figure 3b shows the MNAR situation when the missing variable is d-connected to its missingness mechanism through an unobserved variable $U$. Finally in figure 3c, MNAR holds because even when conditioning on A to get $X$ conditionally independent of its missingness mechanism (MAR situation), $A$ is not fully observed, and hence MAR does not hold. In MNAR situation (a) and (b), $X$ cannot be recovered without bias while in situation (c) $X$ could be recovered with some residual bias.

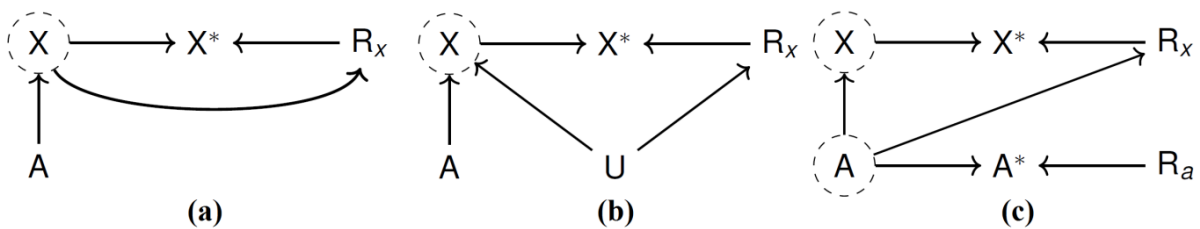

**Figure 3:** Three usual MNAR models. Missing data can be MNAR when there is a (un)directed path from a variable with missing values to its missingness mechanism (a), or through an unobserved variable U (b) or MNAR holds when all variables of a model are partially observed (c). Nodes with dashed circle represent variables that would have been observed had they not missing values. Nodes with a star represent observed variables with missing values.

The missingness graphs in figure 4 shows situations that can be found when analysing repeated immunological biomarkers. In figure 4a, $X_{i,t=1}$ is d-connected to its missingness

mechanism $R_{x_i,t=1}$. However, when conditioning on $X_{i,t=0}, X_{j,t=0}, X_{j,t=1}$, $X_{i,t=1}$ becomes conditionally independent of its missingness mechanism: $R_{x_i,t=1} \perp X_{i,t=1}|X_{i,t=0}, X_{j,t=0}, X_{j,t=1}$. Thus, in this case, MAR holds and the joint distribution $P(X_{i,t=0}, X_{j,t=0}, X_{i,t=1}, X_{j,t=1})$ is recoverable using multiple imputation with $X_{i,t=0}, X_{j,t=0}, X_{j,t=1}$ as predictor. Figure 4b illustrates a MNAR situation where a variable used to block a path (used as predictor) is partially observed (cf figure 3c). The d-separation of $X_{i,t=1}$ and $X_{j,t=1}$ from their missingness mechanisms $R_{x_i,t=1}$ and $R_{x_j,t=1}$, requires to condition on $X_{i,t=0}, X_{j,t=0}, X_{j,t=1}$ and $X_{i,t=0}, X_{j,t=0}, X_{i,t=1}$ respectively. Since in both cases, the set of nodes used to d-separate missing variables and their missingness mechanisms are not fully observed, MAR does not hold. Therefore, $X_{i,t=1}$ and $X_{j,t=1}$ can only be recovered with some residual bias in Figure 4b.

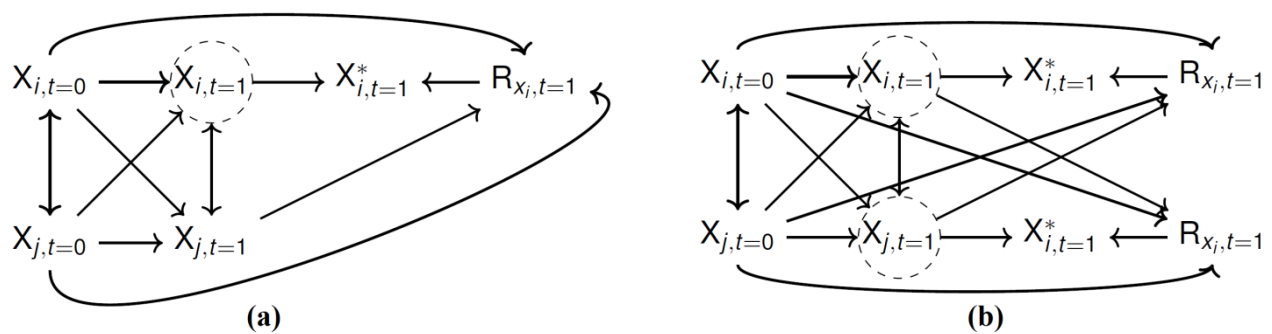

**Figure 4:** Different missingness mechanisms found in immunological biomarkers such as MAR (a) and MNAR (b). Nodes with dashed circle represent variables that would have been observed had they not missing values. Nodes with a star represent observed variables with missing values..

# Appendix E Estimation of the Mean squared error (MSE) of biomarkers with a true effect

The following tables show the mean squared of the biomarkers that have a true effect on the binary outcome in all scenarios tested.

**Table 1:** Mean squared error (MSE) of biomarkers with a true effect. Parameters of the scenario were $p$=20, $n_{visits}$=4, $n_{obs}$=1000 and alpha=0.02

|  | MSE PC-stable | MSE COPC-stable |
| --- | --- | --- |
| **BM6v1** | 0.54 | 0.55 |
| **BM11v1** | 0.80 | 0.80 |
| **BM2v2** | 0.60 | 0.55 |
| **BM4v2** | 0.28 | 0.11 |
| **BM5v2** | 0.65 | 0.76 |
| **BM11v2** | 0.54 | 0.29 |
| **BM12v2** | 0.34 | 0.21 |
| **BM15v2** | 0.69 | 0.45 |
| **BM17v2** | 0.28 | 0.33 |
| **BM18v2** | 0.35 | 0.37 |
| **BM19v2** | 0.68 | 0.82 |
| **BM1v4** | 0.47 | 0.14 |
| **BM2v4** | 0.28 | 0.11 |
| **BM11v4** | 0.29 | 0.16 |
| **BM12v4** | 0.63 | 0.34 |
| **BM14v4** | 0.77 | 0.33 |

**Table 2:** Mean squared error (MSE) of biomarkers with a true effect. Parameters of the scenario were $p$=20, $n_{visits}$=4, $n_{obs}$=1000 and alpha=0.2

|  | MSE PC-stable | MSE COPC-stable |
| --- | --- | --- |
| **BM6v1** | 0.54 | 0.55 |
| **BM11v1** | 0.80 | 0.80 |
| **BM2v2** | 0.60 | 0.51 |
| **BM4v2** | 0.27 | 0.11 |

| | | |
|---|---|---|
| **BM5v2** | 0.74 | 0.71 |
| **BM11v2** | 0.44 | 0.32 |
| **BM12v2** | 0.33 | 0.21 |
| **BM15v2** | 0.71 | 0.41 |
| **BM17v2** | 0.27 | 0.41 |
| **BM18v2** | 0.28 | 0.40 |
| **BM19v2** | 0.81 | 0.79 |
| **BM1v4** | 0.40 | 0.16 |
| **BM2v4** | 0.23 | 0.09 |
| **BM11v4** | 0.23 | 0.12 |
| **BM12v4** | 0.59 | 0.34 |
| **BM14v4** | 0.82 | 0.28 |

**Table 3:** Mean squared error (MSE) of biomarkers with a true effect. Parameters of the scenario were $p=20$, $n_{visits}=4$, $n_{obs}=50$ and alpha=0.02

| | MSE PC-stable | MSE COPC-stable |
|---|---|---|
| **BM6v1** | 0.54 | 0.55 |
| **BM11v1** | 0.80 | 0.80 |
| **BM2v2** | 0.60 | 0.41 |
| **BM4v2** | 0.30 | 0.06 |
| **BM5v2** | 0.81 | 0.62 |
| **BM11v2** | 0.60 | 0.47 |
| **BM12v2** | 0.36 | 0.11 |
| **BM15v2** | 0.76 | 0.36 |
| **BM17v2** | 0.42 | 0.29 |
| **BM18v2** | 0.54 | 0.32 |
| **BM19v2** | 0.98 | 0.82 |
| **BM1v4** | 0.50 | 0.09 |
| **BM2v4** | 0.38 | 0.13 |
| **BM11v4** | 0.36 | 0.07 |
| **BM12v4** | 0.70 | 0.12 |
| **BM14v4** | 0.97 | 0.21 |

**Table 4:** Mean squared error (MSE) of biomarkers with a true effect. Parameters of the scenario were $p=20$, $n_{visits}=4$, $n_{obs}=50$ and alpha=0.2

|         | MSE PC-stable | MSE COPC-stable |
|---------|---------------|-----------------|
| **BM6v1**  | 0.47 | 0.55 |
| **BM11v1** | 0.77 | 0.80 |
| **BM2v2**  | 0.55 | 0.20 |
| **BM4v2**  | 0.29 | 0.06 |
| **BM5v2**  | 0.75 | 0.39 |
| **BM11v2** | 0.52 | 0.15 |
| **BM12v2** | 0.32 | 0.08 |
| **BM15v2** | 0.70 | 0.24 |
| **BM17v2** | 0.40 | 0.14 |
| **BM18v2** | 0.48 | 0.16 |
| **BM19v2** | 0.92 | 0.43 |
| **BM1v4**  | 0.43 | 0.06 |
| **BM2v4**  | 0.35 | 0.08 |
| **BM11v4** | 0.31 | 0.05 |
| **BM12v4** | 0.67 | 0.09 |
| **BM14v4** | 0.90 | 0.16 |

**Table 5:** Mean squared error (MSE) of biomarkers with a true effect. Parameters of the scenario were $p=20$, $n_{visits}=6$, $n_{obs}=1000$ and alpha=0.02

|         | MSE PC-stable | MSE COPC-stable |
|---------|---------------|-----------------|
| **BM9v1**  | 0.20 | 0.27 |
| **BM11v1** | 0.30 | 0.36 |
| **BM15v1** | 0.67 | 0.75 |
| **BM3v2**  | 0.31 | 0.21 |
| **BM5v2**  | 0.65 | 0.60 |
| **BM11v2** | 0.90 | 0.46 |
| **BM12v2** | 0.55 | 0.84 |
| **BM17v2** | 0.41 | 0.54 |

|  | | |
|---|---|---|
| **BM18v2** | 0.60 | 0.73 |
| **BM2v3**  | 0.53 | 0.22 |
| **BM3v3**  | 0.76 | 0.52 |
| **BM4v3**  | 0.22 | 0.15 |
| **BM9v3**  | 0.55 | 0.33 |
| **BM19v3** | 0.21 | 0.38 |
| **BM4v4**  | 0.40 | 0.12 |
| **BM7v4**  | 0.42 | 0.52 |
| **BM8v4**  | 0.91 | 0.51 |
| **BM9v4**  | 0.61 | 0.65 |
| **BM16v4** | 0.32 | 0.17 |
| **BM6v5**  | 0.60 | 0.31 |
| **BM15v5** | 0.55 | 0.24 |
| **BM7v6**  | 0.94 | 0.45 |
| **BM8v6**  | 0.34 | 0.10 |

**Table 6:** Mean squared error (MSE) of biomarkers with a true effect. Parameters of the scenario were $p$=20, $n_{visits}$=6, $n_{obs}$=1000 and alpha=0.2

|  | MSE PC-stable | MSE COPC-stable |
|---|---|---|
| **BM9v1**  | 0.23 | 0.27 |
| **BM11v1** | 0.26 | 0.36 |
| **BM15v1** | 0.74 | 0.75 |
| **BM3v2**  | 0.29 | 0.22 |
| **BM5v2**  | 0.54 | 0.64 |
| **BM11v2** | 0.83 | 0.60 |
| **BM12v2** | 0.70 | 0.77 |
| **BM17v2** | 0.41 | 0.47 |
| **BM18v2** | 0.44 | 0.70 |
| **BM2v3**  | 0.45 | 0.17 |
| **BM3v3**  | 0.51 | 0.44 |
| **BM4v3**  | 0.18 | 0.15 |
| **BM9v3**  | 0.49 | 0.25 |
| **BM19v3** | 0.21 | 0.35 |
| **BM4v4**  | 0.36 | 0.18 |
| **BM7v4**  | 0.39 | 0.48 |

| | | |
|---|---|---|
| **BM8v4** | 0.85 | 0.46 |
| **BM9v4** | 0.50 | 0.56 |
| **BM16v4** | 0.26 | 0.15 |
| **BM6v5** | 0.49 | 0.29 |
| **BM15v5** | 0.40 | 0.21 |
| **BM7v6** | 0.63 | 0.36 |
| **BM8v6** | 0.26 | 0.09 |
| **BM9v6** | 0.45 | 0.43 |

**Table 7:** Mean squared error (MSE) of biomarkers with a true effect. Parameters of the scenario were $p=20$, $n_{visits}=6$, $n_{obs}=50$ and alpha=0.02

| | MSE PC-stable | MSE COPC-stable |
|---|---|---|
| **BM9v1** | 0.26 | 0.27 |
| **BM11v1** | 0.35 | 0.36 |
| **BM15v1** | 0.75 | 0.75 |
| **BM3v2** | 0.31 | 0.29 |
| **BM5v2** | 0.71 | 0.24 |
| **BM11v2** | 0.95 | 0.47 |
| **BM12v2** | 0.93 | 0.82 |
| **BM17v2** | 0.63 | 0.58 |
| **BM18v2** | 0.74 | 0.63 |
| **BM2v3** | 0.55 | 0.34 |
| **BM3v3** | 0.77 | 0.19 |
| **BM4v3** | 0.24 | 0.15 |
| **BM9v3** | 0.60 | 0.29 |
| **BM19v3** | 0.53 | 0.22 |
| **BM4v4** | 0.43 | 0.26 |
| **BM7v4** | 0.83 | 0.39 |
| **BM8v4** | 0.96 | 0.54 |
| **BM9v4** | 0.68 | 0.36 |
| **BM16v4** | 0.37 | 0.10 |
| **BM6v5** | 0.82 | 0.22 |
| **BM15v5** | 0.73 | 0.12 |
| **BM7v6** | 0.97 | 0.25 |
| **BM8v6** | 0.36 | 0.05 |
| **BM9v6** | 0.87 | 0.26 |

**Table 8:** Mean squared error (MSE) of biomarkers with a true effect. Parameters of the scenario were $p=20$, $n_{visits}=6$, $n_{obs}=50$ and alpha=0.2

|         | MSE PC-stable | MSE COPC-stable |
|---------|---------------|-----------------|
| **BM9v1**  | 0.23 | 0.27 |
| **BM11v1** | 0.30 | 0.36 |
| **BM15v1** | 0.71 | 0.75 |
| **BM3v2**  | 0.30 | 0.21 |
| **BM5v2**  | 0.65 | 0.19 |
| **BM11v2** | 0.84 | 0.36 |
| **BM12v2** | 0.90 | 0.59 |
| **BM17v2** | 0.55 | 0.34 |
| **BM18v2** | 0.67 | 0.35 |
| **BM2v3**  | 0.53 | 0.13 |
| **BM3v3**  | 0.70 | 0.13 |
| **BM4v3**  | 0.23 | 0.08 |
| **BM9v3**  | 0.55 | 0.16 |
| **BM19v3** | 0.49 | 0.20 |
| **BM4v4**  | 0.40 | 0.11 |
| **BM7v4**  | 0.78 | 0.31 |
| **BM8v4**  | 0.87 | 0.36 |
| **BM9v4**  | 0.67 | 0.31 |
| **BM16v4** | 0.35 | 0.06 |
| **BM6v5**  | 0.76 | 0.17 |
| **BM15v5** | 0.71 | 0.11 |
| **BM7v6**  | 0.93 | 0.26 |
| **BM8v6**  | 0.32 | 0.05 |
| **BM9v6**  | 0.78 | 0.15 |

# Appendix F Estimation of biomarkers' median effect and PCER (per-comparison error rate)

**Table 1:** Immunological biomarkers with a PCER < 0.5% in model 2. The number following "v" in each biomarker's name stands for the visit number. See additional file 3 for the complete description of the biomarkers.

| | Death | | | Progression | | | Toxicity | | |
|---|---|---|---|---|---|---|---|---|---|
| Rank | Biomarker | Median effect | PCER | Biomarker | Median effect | PCER | Biomarker | Median effect | PCER |
| 1 | BM16v2 | 0.81 | 0.0035 | BM8v1 | 0.77 | 0.0031 | BM7v4 | 0.79 | 0.0028 |
| 2 | BM5v1 | 0.81 | 0.0035 | BM44v4 | 0.72 | 0.0036 | BM8v4 | 0.76 | 0.0034 |
| 3 | BM42v3 | 0.80 | 0.0037 | BM26v2 | 0.71 | 0.0041 | BM16v3 | 0.75 | 0.0036 |
| 4 | BM48v1 | 0.86 | 0.0037 | BM30v3 | 0.71 | 0.0041 | BM26v4 | 0.75 | 0.0036 |
| 5 | BM42v2 | 0.80 | 0.0038 | BM44v3 | 0.68 | 0.0042 | BM7v3 | 0.76 | 0.0037 |
| 6 | BM14v4 | 0.79 | 0.0039 | BM45v1 | 0.70 | 0.0047 | BM9v4 | 0.72 | 0.0039 |
| 7 | BM30v4 | 0.80 | 0.0039 | BM39v4 | 0.66 | 0.0049 | BM39v3 | 0.72 | 0.0039 |
| 8 | BM11v4 | 0.83 | 0.0040 | BM40v4 | 0.66 | 0.0049 | BM32v3 | 0.71 | 0.0042 |
| 9 | BM11v1 | 0.76 | 0.0042 | BM14v2 | 0.66 | 0.0050 | BM30v1 | 0.67 | 0.0045 |
| 10 | BM9v4 | 0.81 | 0.0043 | | | | BM45v1 | 0.71 | 0.0046 |
| 11 | BM17v1 | 0.77 | 0.0043 | | | | BM18v3 | 0.71 | 0.0048 |
| 12 | BM11v2 | 0.78 | 0.0045 | | | | | | |
| 13 | BM30v1 | 0.76 | 0.0046 | | | | | | |
| 14 | BM31v3 | 0.75 | 0.0047 | | | | | | |
| 15 | BM48v3 | 0.78 | 0.0048 | | | | | | |
| 16 | BM25v2 | 0.77 | 0.0049 | | | | | | |
| 17 | BM10v4 | 0.77 | 0.0050 | | | | | | |

**Table 2:** Immunological biomarkers with a PCER < 0.5% in model 3. The number following "v" in each biomarker's name stands for the visit number. See additional file 3 for the complete description of the biomarkers.

| Rank | Death Biomarker | Median effect | PCER | Progression Biomarker | Median effect | PCER | Toxicity Biomarker | Median effect | PCER |
|---|---|---|---|---|---|---|---|---|---|
| 1 | BM60v3 | 0.87 | 0.0015 | BM78v4 | 0.74 | 0.0024 | BM96v4 | 0.86 | 0.0013 |
| 2 | BM87v1 | 0.83 | 0.0021 | BM58v2 | 0.74 | 0.0025 | BM87v1 | 0.81 | 0.0018 |
| 3 | BM96v4 | 0.84 | 0.0021 | BM108v4 | 0.72 | 0.0025 | BM8v4 | 0.80 | 0.0019 |
| 4 | BM105v4 | 0.83 | 0.0022 | BM105v4 | 0.73 | 0.0026 | BM9v4 | 0.72 | 0.0026 |
| 5 | BM58v2 | 0.81 | 0.0025 | BM8v1 | 0.72 | 0.0028 | BM7v4 | 0.76 | 0.0029 |
| 6 | BM58v4 | 0.80 | 0.0026 | BM108v3 | 0.68 | 0.0031 | BM57v3 | 0.75 | 0.0029 |
| 7 | BM11v4 | 0.81 | 0.0026 | BM86v4 | 0.63 | 0.0031 | BM88v3 | 0.69 | 0.0032 |
| 8 | BM76v3 | 0.80 | 0.0027 | BM80v1 | 0.71 | 0.0031 | BM59v1 | 0.76 | 0.0032 |
| 9 | BM108v4 | 0.80 | 0.0027 | BM59v4 | 0.71 | 0.0032 | BM62v3 | 0.72 | 0.0033 |
| 10 | BM11v1 | 0.78 | 0.0027 | BM89v3 | 0.67 | 0.0032 | BM93v3 | 0.72 | 0.0034 |
| 11 | BM9v4 | 0.80 | 0.0028 | BM65v4 | 0.69 | 0.0032 | BM65v3 | 0.69 | 0.0034 |
| 12 | BM68v4 | 0.77 | 0.0028 | BM66v4 | 0.68 | 0.0032 | BM89v3 | 0.72 | 0.0034 |
| 13 | BM98v4 | 0.78 | 0.0028 | BM56v3 | 0.68 | 0.0033 | BM7v3 | 0.72 | 0.0036 |
| 14 | BM52v1 | 0.78 | 0.0028 | BM86v3 | 0.68 | 0.0033 | BM63v1 | 0.67 | 0.0037 |
| 15 | BM87v2 | 0.81 | 0.0028 | BM63v1 | 0.68 | 0.0034 | BM69v3 | 0.65 | 0.0038 |
| 16 | BM63v4 | 0.77 | 0.0029 | BM67v3 | 0.69 | 0.0035 | BM11v2 | 0.68 | 0.0040 |
| 17 | BM74v4 | 0.77 | 0.0029 | BM70v3 | 0.68 | 0.0036 | BM104v3 | 0.62 | 0.0040 |
| 18 | BM11v2 | 0.78 | 0.0030 | BM83v3 | 0.65 | 0.0039 | BM6v2 | 0.67 | 0.0041 |
| 19 | BM59v1 | 0.76 | 0.0031 | BM9v2 | 0.63 | 0.0039 | BM80v3 | 0.69 | 0.0043 |
| 20 | BM63v3 | 0.77 | 0.0032 | BM56v4 | 0.65 | 0.0041 | BM103v3 | 0.66 | 0.0043 |
| 21 | BM65v3 | 0.75 | 0.0033 | BM96v2 | 0.62 | 0.0043 | BM10v2 | 0.64 | 0.0043 |
| 22 | BM100v3 | 0.76 | 0.0033 | BM61v3 | 0.63 | 0.0044 | BM98v2 | 0.65 | 0.0046 |
| 23 | BM10v4 | 0.77 | 0.0033 | BM60v4 | 0.63 | 0.0045 | BM96v3 | 0.69 | 0.0046 |
| 24 | BM101v3 | 0.75 | 0.0034 | BM98v1 | 0.61 | 0.0046 | BM83v3 | 0.65 | 0.0046 |
| 25 | BM87v3 | 0.69 | 0.0036 | BM101v3 | 0.65 | 0.0046 | BM75v3 | 0.45 | 0.0047 |
| 26 | BM105v2 | 0.71 | 0.0037 | BM56v2 | 0.64 | 0.0047 | BM91v2 | 0.60 | 0.0047 |
| 27 | BM5v1 | 0.76 | 0.0037 | BM98v2 | 0.61 | 0.0047 | BM72v3 | 0.62 | 0.0048 |
| 28 | BM106v4 | 0.75 | 0.0038 | BM10v3 | 0.61 | 0.0047 | BM90v3 | 0.60 | 0.0049 |
| 29 | BM6v3 | 0.70 | 0.0039 | BM55v4 | 0.59 | 0.0048 | BM56v2 | 0.65 | 0.0049 |
| 30 | BM57v3 | 0.69 | 0.0039 | BM93v4 | 0.62 | 0.0050 | BM9v3 | 0.64 | 0.0050 |
| 31 | BM54v4 | 0.71 | 0.0040 | BM100v4 | 0.65 | 0.0050 | | | |
| 32 | BM7v2 | 0.72 | 0.0040 | BM85v2 | 0.61 | 0.0050 | | | |
| 33 | BM79v3 | 0.71 | 0.0040 | | | | | | |
| 34 | BM8v2 | 0.63 | 0.0043 | | | | | | |
| 35 | BM70v3 | 0.69 | 0.0045 | | | | | | |
| 36 | BM61v2 | 0.74 | 0.0047 | | | | | | |
| 37 | BM65v4 | 0.65 | 0.0048 | | | | | | |
| 38 | BM78v4 | 0.75 | 0.0050 | | | | | | |
| 39 | BM63v1 | 0.71 | 0.0050 | | | | | | |


 \end{appendices}
\end{document}